\documentclass[]{jfm}

\usepackage[]{amsmath}
\usepackage[]{xcolor}
\usepackage[]{subfigure}
\usepackage{upgreek}
\usepackage{overpic}
\usepackage[]{multirow}
\usepackage[]{url}
\usepackage[]{nicefrac}
\usepackage{bm}
\usepackage[]{graphicx}
\usepackage{booktabs}
\usepackage{float}
\usepackage{tabularx}
\usepackage{pgf,tikz}
\usepackage{pgfplots}

\graphicspath{{Figures/}}

\begin{document}

\newtheorem{lemma}{Lemma}
\newtheorem{corollary}{Corollary}

\shorttitle{Metric for attractor overlap} 
\shortauthor{R.~Ishar and others} 

\title{Metric for attractor overlap}

\author{Rishabh Ishar\aff{1}\corresp{\email{rishabhishar.bemech14@pec.edu.in}}\footnote{Present address : Department of Mechanical Engineering, Massachusetts Institute of \\ \phantom \qquad \qquad \qquad Technology, 77 Massachusetts Avenue, Cambridge, MA 02139, USA},
Eurika Kaiser\aff{2}, Marek Morzynski\aff{3}, Daniel Fernex\aff{4}, Richard Semaan\aff{4}, Marian Albers\aff{5}, Pascal S. Meysonnat\aff{5}, Wolfgang Schr\"{o}der\aff{5,6}, and Bernd R.~Noack\aff{7,4,8,9}}

\affiliation
{
\aff{1}
Department of Mechanical Engineering, Punjab Engineering College, Chandigarh 160012, India
\aff{2}
University  of Washington, 
Department of  Mechanical Engineering,
Stevens Way, Box 352600,
Seattle, WA 98195, United States
\aff{3}
Pozna\'n University of Technology,
Chair of Virtual Engineering, 
Jana Pawla II 24, PL 60-965 Pozna\'n, Poland
\aff{4}
Institut f\"{u}r Str\"{o}mungsmechanik, Technische Universit\"{a}t Braunschweig, Hermann-Blenk-Str. 37,
38108 Braunschweig, Germany
\aff{5}
Institute of Aerodynamics, RWTH Aachen University, W\"{u}llnerstr. 5a, 52062 Aachen, Germany
\aff{6}
Forschungszentrum J\"{u}lich, JARA-High-Performance Computing, 52425 J\"{u}lich, Germany
\aff{7}
Laboratoire d'Informatique pour la M\'ecanique et les Sciences de l'Ing\'enieur,
LIMSI-CNRS,  Rue John von Neumann, Campus Universitaire d'Orsay, B\^at 508,
91403 Orsay Cedex, France
\aff{8}
Institut f\"ur Str\"omungsmechanik und Technische Akustik (ISTA),
Technische Universit\"at Berlin,
M\"uller-Breslau-Stra{\ss}e 8,
10623 Berlin, Germany
\aff{9}
Institute for Turbulence-Noise-Vibration Interaction and Control,
Harbin Institute of Technology, Shenzhen Campus,  China
}

\maketitle

\begin{abstract}
We present the first general metric for attractor overlap (MAO) 
facilitating an unsupervised comparison of flow data sets.
The starting point is two or more attractors, 
i.e., ensembles of states representing different operating conditions.
The proposed metric generalizes the standard Hilbert-space distance
between two snapshots to snapshot ensembles of two attractors.
A reduced-order analysis for big data and many attractors
is enabled by coarse-graining the snapshots 
into  representative clusters
with corresponding centroids and population probabilities.
For a large number of attractors,
MAO is augmented by proximity maps for the snapshots, the centroids, and the attractors,
giving scientifically interpretable visual access to the closeness of the states.
The coherent structures
belonging to the overlap and disjoint states between these attractors 
are distilled by few representative centroids.We employ MAO for two quite different actuated flow configurations:
a two-dimensional wake with vortices in a narrow frequency range
and three-dimensional wall turbulence with broadband spectrum.
In the first application,
seven control laws are applied to the fluidic pinball,
i.e., the two-dimensional flow around three circular cylinders whose centers form an equilateral triangle pointing in the upstream direction.
These seven operating conditions comprise 
unforced shedding, boat tailing, base bleed, high- and low-frequency forcing
as well as two opposing Magnus effects.
In the second example, 
MAO is applied to three-dimensional simulation data 
from an open-loop drag reduction study of a turbulent boundary layer.
The actuation mechanisms of 38 spanwise traveling transversal surface waves are investigated.
MAO compares and classifies these actuated flows in agreement with physical intuition.
For instance, the first feature coordinate of the attractor proximity map 
correlates with drag for the fluidic pinball and for the turbulent boundary layer.
MAO has a large spectrum of potential applications
ranging from   a quantitative comparison between numerical simulations and experimental particle-image velocimetry data
to the analysis of simulations representing a myriad of different operating conditions.
\end{abstract}
\renewcommand{\vec}[1]{\bm{#1}}
\section{Introduction}
\label{Sec:Introduction}

In this study, 
we propose arguably the first general metric 
between attractor data 
from different operating conditions.
With \emph{attractor data}, 
we refer to an ensemble of statistically representative flow snapshots
which allow for the computation of statistical moments 
and resolve coarse-grained coherent structures.
Here, attractor is understood in a nonlinear dynamics sense for dissipative systems,
i.e., a subset of the state space 
to which all solutions converge independently of the initial condition 
\citep[see, e.g.,][]{Schuster1988book}.
The existence of a single global attractor is implicitly assumed
in statistical fluid mechanics.
Otherwise,  statistical moments may have multiple values 
depending on the initial conditions.
The focus on attractor data is not requested by the metric
but simplifies the first demonstration of its usefulness.

In particular,
we propose an unsupervised comparison methodology  with little subjective  bias.
This methodology is exemplified for two configurations 
with associated many open-loop actuations each.
The first example is the two-dimensional \emph{fluidic pinball},
i.e., the flow around three stationary rotating circular cylinders \citep{Noack2016jfm}.
Unlike the single rotating cylinder, 
the  fluidic pinball exhibits rich spatial-temporal dynamics under different actuation laws
at similarly low computational cost \citep{Noack2017put}.
The second example demonstrates the applicability to three-dimensional wall-bounded turbulent flow,
namely the turbulent boundary layer actuated 
with a spanwise transversely traveling surface waves \citep{Meysonnat2016}.

Most fluid mechanics publications contain a comparison of flows 
from different sources,
e.g.\ experiments versus simulations 
or at various operating conditions, 
e.g.\ optimal control versus the unforced benchmark.

In an engineering application, this comparison is easily performed
for a single global solution parameter of interest, e.g.\ drag for a car,
lift for an airfoil, mixing for a combustor or far-field noise of a jet engine.

These comparisons of performance parameters are simple 
but provide  a limited assessment of the flow physics.
For instance,
Reynolds-averaged Navier-Stokes (RANS) simulations of cylinder wakes 
may predict well the drag coefficient at low Reynolds number.
However, RANS computations predict the von K\'arm\'an vortex street 
to dissipate far too quickly in streamwise direction.
A commonly used and more refined comparison includes 
the statistical moments of the flow field,
or at least transverse or streamwise velocity profiles.
The comparison of statistical moments is straightforward 
using a corresponding Hilbert space norm.
This comparison is more detailed than employing a single order parameter. 
Yet, it excludes spatial-temporal dynamics of coherent structures.

Coherent structures are often visible to the naked eye,
as beautifully depicted for wakes over 500 years ago by Leonardo da Vinci.
Their quantification has been subject 
of thousands of publications and many disputes.
Vortices provide important dynamical insight 
into geometrically simple two- and three-dimensional flows
and have been the cornerstone of early reduced-order modeling efforts
starting with the famous Helmholtz vortex laws in 1869~\citep[see, e.g.][]{Lugt1996book}.
Data-driven vortex identifications have been proposed by \citet{Jeong1995jfm} 
for snapshots
and by \citet{Haller2005jfm} for flow histories. 
These frequently cited  publications
represent---pars pro toto---myriad of other feature analyses,
e.g.\ Galilean invariant snapshot topology \citep{Kasten2016am}. 

An alternative approach is a reduced-order representation by expansions in terms of global modes.
Proper orthogonal decomposition \citep{Berkooz1993arfm},
dynamic mode decomposition \citep{Rowley2009jfm,Schmid2010jfm},
stability eigenmodes \citep{Theofilis2011arfm} 
or plain temporal Fourier expansions may serve as examples.
These modes provide important insight into  physical mechanisms of the coherent structure dynamics.
Yet, one cannot expect the modes of different sources 
to coincide or even to be similar.
In addition, the energy resolved by low-order turbulence representations 
is often much smaller than the energy of the unresolved stochastic fluctuations.


The analyses of local features and expansions in terms of global modes pose significant challenges
for an automated comparison of different attractor data.
In this study, 
we neither follow the Galerkin method nor the vortex modeling approach.
Instead, the distance between two data sets is, roughly speaking,
 geometrically characterized 
by the average distance between each snapshot of one attractor
to the closest snapshot of the other attractor.
In characterizing the overlap and disjoint regions of both attractors
by selected flow states,
we follow the pioneering clustering approach of \cite{Burkardt2006cmame} in fluid dynamics.
Clustering implies that similar snapshots are put into `bins' represented by a centroid.
The centroid can be understood as a flow state  averaged over all elements in the same `bin'.
Shared centroids span the overlap region of both attractors
 and disjoint centroids illustrate different attractor regions.
The comparison methodology is augmented by powerful feature extraction from machine learning.
The beauty and ingredients of this approach
will be elucidated in the following section.

For the proposed framework, 
we choose a direct numerical simulation of the fluidic pinball, 
i.e., the flow around three equal circular cylinders
with centers on an equilateral triangle pointing in upstream direction \citep{Noack2016jfm}.
The term `fluidic pinball' is owed to the possibility of moving fluid particles 
like balls in a conventional pinball machine 
by suitably rotating the cylinders. 
The pinball configuration includes most wake stabilization strategies
with suitable rotation of the three cylinders.
Examples include   phasor control \citep{Roussopoulos1993jfm},
aerodynamic boat tailing  \citep{Geropp1995patent,Geropp2000ef,Barros2016jfm},
base bleed \citep{Wood1964jras,Bearman1967aq},
high-frequency forcing \citep{Thiria2006jfm,Oxlade2015jfm} and
low-frequency forcing \citep{Pastoor2008jfm}.
Another possibility is to deflect the wake via the Magnus effect.
In this study, we compare the unforced reference and six open-loop actuation mechanisms.

Wakes or free-shear flows downstream of blunt bodies often possess dominant structures
as opposed to wall-bounded shear flows 
whose distributions of the spatial and temporal scales exhibit smooth broadband spectra. 
As a challenging benchmark, the comparison methodology is also applied to data from a large-eddy 
simulation of a zero pressure-gradient boundary layer over a surface that undergoes a transversal spanwise traveling wave motion, i.e., the wall is deflected in the wall-normal direction. 
The canonical fully turbulent boundary layer flow is an interesting test case
to study the impact of wall motion on friction drag. 
The results of this simple geometry are to a certain extent transferable to problems 
such as airfoil flows where the boundary layer varies in the streamwise direction. 
Besides the well-known passive control approaches,
 e.g., riblets \citep{Garcia-Mayoral2011}, 
a wide range of active drag reduction methods has been investigated in the past three decades. 
To name a few, \cite{Jung1992} achieved high relative drag reduction
using in-plane spanwise wall oscillations, \cite{Du2002} applied a traveling 
wave-like body force in the spanwise direction to lower the friction drag, 
and \cite{Zhao2004} extended the idea by an flexible wall approach. 
A good overview of active
drag reduction approaches is given by \cite{Quadrio2011}.

Spanwise traveling transversal surface waves have been investigated experimentally 
\citep{Itoh2006,Tamano2012,Li2015} and numerically \citep{Klumpp2010b,Koh2015a,Meysonnat2016}.
Drag reductions of the order of $10\> \%$ were achieved. 
The physical mechanism  of this active control
is the generation of a secondary near-wall flow field in the wall-normal 
and spanwise direction through a, for example, sinusoidal up- and down 
motion of the wall to interrupt the near-wall cycle of the turbulent shear flow 
and as such to redistribute the turbulent scales. 
The main parameters for the sinusoidal wave actuation are  wavelength, amplitude, 
and frequency. It goes without saying that due to the non-linear interaction between the wall-motion parameters and the wall-shear stress, i.e., the friction drag, it is quite a challenge to efficiently determine for a given flow, i.e., predefined freestream Reynolds number, the optimum parameter settings to minimize the wall-shear stress distribution.
In this study, the large-eddy simulation data for a non-actuated turbulent boundary layer flow constitutes the reference problem.
The drag reduction is studied for 37  transverse surface wave actuations 
which vary in wavelength, amplitude, and frequency.

The structure of the manuscript is as follows.
In \S~\ref{Sec:Plant}, the employed fluidic pinball simulation and the actuated turbulent 
boundary layer flow are described.
The corresponding data includes converged data of the 7 attractors and of the 38 parameter
variations of the traveling wave. 
In \S~\ref{Sec:Method}, the comparison methodology for attractor data is outlined.
The proposed approach is exemplified 
for all operating conditions of the fluidic pinball  in \S~\ref{Sec:Pinball}
and for all actuations of the turbulent boundary layer in \S~\ref{Sec:TBL:Plant}.
Section \S~\ref{Sec:Conclusions} summarizes this study and outlines future directions of research. 
\section{Plant configurations}
\label{Sec:Plant}

The comparison methodology MAO is applied to two configurations
with many data sets from open-loop actuation studies.
A two-dimensional fluidic pinball configuration  (\S~\ref{Sec:Pinball:Plant})
is computationally inexpensive and allows an extensive visualization of the flow quantities
while exhibiting a complex behavior.
The second set from an open-loop drag reduction  study
of a turbulent boundary layer (\S~\ref{Sec:TBL:Plant}) 
shall allow to assess MAO for complex three-dimensional flow data
with a large range of scales and frequencies. 

\subsection{Fluidic pinball simulation}
\label{Sec:Pinball:Plant}

In this section, 
the computation of the employed flow data is described.
In \S~\ref{Sec:Pinball:Configuration},
the configuration of the fluidic pinball is introduced.
The corresponding direct Navier-Stokes solver is described in \S~\ref{Sec:Pinball:DNS}. 
Finally, the fluidic pinball simulation 
of seven  open-loop actuations is detailed.
These flow data are subjected to the proposed comparison methodology of the next section.

\subsubsection{Fluidic pinball configuration}
\label{Sec:Pinball:Configuration}

\begin{figure}
	\centering
	\includegraphics[height=4.5cm]{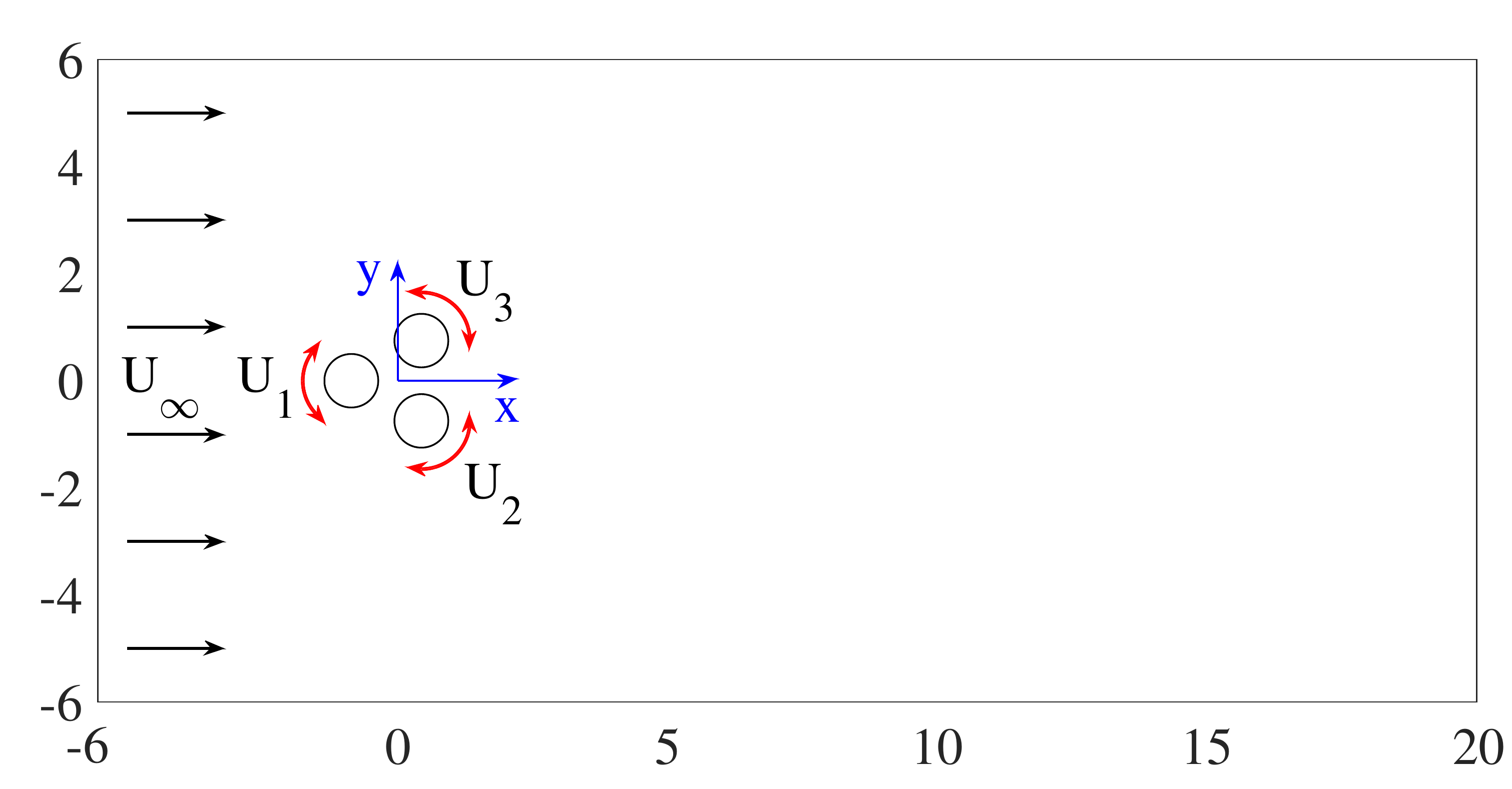}
	\caption{Configuration of the fluidic pinball.
         The Cartesian coordinate system $(x,y)$ is depicted at the center of the cylinders.}
	\label{Fig:Pinball:Configuration}
\end{figure}

In the following,
the considered two-dimensional flow control configuration is described.
Three equal circular cylinders with radius $R$ 
are placed parallel to each other in a viscous incompressible uniform flow 
at speed $U_{\infty}$ (see Fig.\ \ref{Fig:Pinball:Configuration}).
The centers of the cylinders form an equilateral triangle
with side length $3R$, symmetrically positioned with respect to the flow. 
The leftmost triangle vertex points upstream, 
while the rightmost side is orthogonal to the oncoming flow.
Thus, the transverse extent of the three-cylinder configuration
is given by $5R$.

This flow is described in a Cartesian coordinate system 
where the $x$-axis points in the direction of the flow,
the $z$-axis is aligned with the cylinder axes, 
and the $y$-axis is orthogonal to both 
(see figure \ref{Fig:Pinball:Configuration}).
The origin $\vec{0}$ of this coordinate system coincides
with the geometric center of the cylinder triangle.
The location is denoted by 
$\vec{x}=(x,y,z)=x  \vec{e}_x + y  \vec{e}_y + z  \vec{e}_z$,
where $\vec{e}_{x,y,z}$ are unit vectors pointing in the direction
of the corresponding axes.
Analogously, the velocity reads
$\vec{u}=(u,v,w)=u  \vec{e}_x + v  \vec{e}_y + w  \vec{e}_z$.
The pressure is denoted  by $p$ and the time by $t$.
In the following, we assume a two-dimensional flow,
i.e., no dependence of any flow quantity on $z$ 
and vanishing spanwise velocity $w \equiv 0$.

The Newtonian fluid is characterized 
by a constant density $\rho$
and kinematic viscosity $\nu$. 
In the following, 
all quantities are assumed to be non-dimensionalized
with cylinder diameter $D=2R$, the velocity $U_{\infty}$
and the fluid density $\rho$.
The corresponding Reynolds number is $Re_D=100$ where $Re_D= U_{\infty} D / \nu$.
The Reynolds number based on the transverse length $L=5D$ is $2.5$ times larger.
The non-dimensionalization with respect to the diameter is more common for clusters of cylinders
\citep{Hu2008jfm,Hu2008jfm2,Bansal2017etfs}
and will be adopted in the following.
With this non-dimensionalization, 
the cylinder axes are located at 
\begin{equation}
\begin{array}{ll@{\quad}ll}
  x_F &= - \sqrt{3}/2 , 
& y_F & = 0,
\\
  x_B & =  \sqrt{3}/4, 
& y_B & = -3/4,
\\
  x_T & =  \sqrt{3}/4,
& y_T & = +3/4.
\\
\end{array}
\end{equation}
Here, and in the following, 
the subscripts `$F$', `$B$' and `$T$' refer to the front, bottom and top cylinder.
An alternate reference is the subscripts $1$, $2$, $3$ 
for the front, bottom, and top cylinder, respectively.
The numbering is in mathematically positive orientation.

The incompressibility condition reads 
\begin{equation}
\label{Eqn:Incompressibility}
\nabla \cdot \vec{u} = 0 ,
\end{equation}
where $\nabla$ represents the Nabla operator.
The evolution is described by the Navier-Stokes equations,
\begin{equation}
\label{Eqn:NavierStokes}
\partial_t \vec{u} + \nabla \cdot \vec{u} \otimes \vec{u}  =   - \nabla p + \frac{1}{Re_D} \triangle \vec{u},
\end{equation}
where $\partial_t$ and $\triangle$ denote the partial derivative with respect to time $t$ and the Laplace operator, respectively.
The dot `$\cdot$' and dyadic product sign `$\otimes$' refer to inner and outer tensor products.

Without forcing, the boundary conditions comprise a no-slip condition 
on the cylinder and a free-stream condition in the far field:
\begin{equation}
\label{Eqn:BoundaryCondition}
\vec{u} = \vec{0} \hbox{\ on the cylinder and }
\vec{u} = \vec{e}_x \hbox{\ at infinity.}
\end{equation}

As initial condition, we chose the unstable steady Navier-Stokes solution $\vec{u}_s(\vec{x})$.
This solution is computed with a Newton search algorithm, like in \citep{Noack2003jfm}.
Vortex shedding is kick-started with cylinder rotations in the first period.

The forcing is exerted by rotation of the cylinders with circumferential velocities
$b_1= U_1= U_F$, $b_2= U_2=U_B$ and $b_3= U_3=U_T$ for the front, bottom and top cylinder, respectively.
The actuation command $\vec{b}=(b_1,b_2,b_3)=(U_1,U_2,U_3)$ 
is preferably used for control theory purposes \citep{Brunton2015amr,Duriez2016book}
while $(U_F,U_B,U_T)$ are more natural for  a discussion of physical mechanisms.  
The actuation is conveniently expressed with the vector cross product `$\times$':
\begin{equation}
\vec{u} = 2 U_i\> \vec{x} \times \vec{e}_z \hbox{\ on the $i$th cylinder.}
\end{equation}
The factor 2 counterbalances  the non-dimensional radius $1/2$.

We like to refer to this configuration as \emph{fluidic pinball} 
as the rotation speeds allow one to change the paths 
of the incoming fluid particles like flippers manipulate the ball of a conventional pinball machine.
The front cylinder rotation may determine if the fluid particle
passes by on the upper or lower side of the cylinder, 
while the top and bottom cylinder may guide the particle through the interior.

\subsubsection{Direct Navier-Stokes solver}
\label{Sec:Pinball:DNS}

\begin{figure}
	\centering
	\includegraphics[height = 4cm]{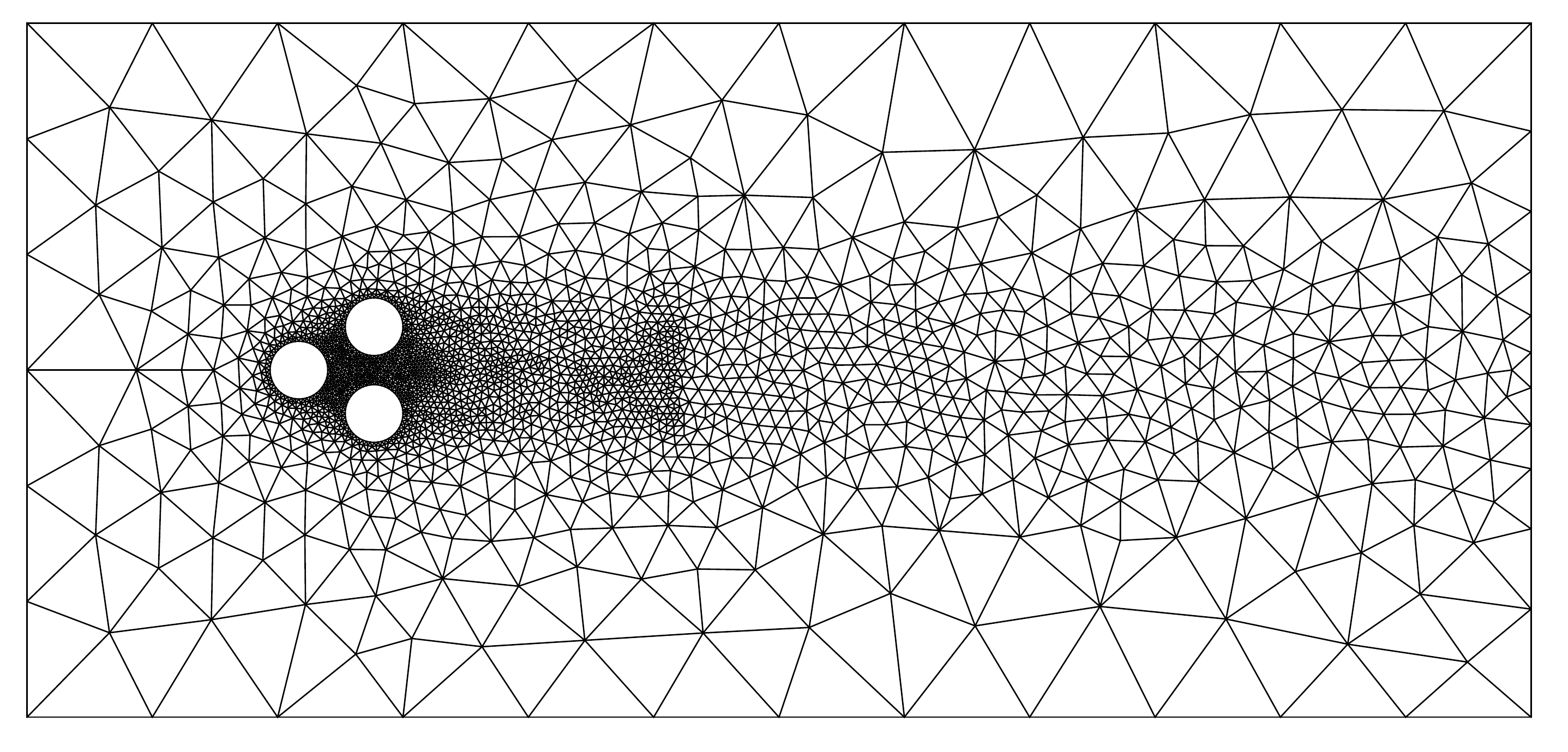}
	\caption{Grid of the fluidic pinball simulation.}
	\label{Fig:Pinball:Grid}
\end{figure}

The chosen computational domain is bounded 
by the rectangle $[-6,20] \times [-6,6]$
and excludes the interior of the cylinders: 
$$ \Omega = \left\{ (x,y) \colon -6 \le x \le 20 \wedge \vert y \vert \le 6  \wedge (x-x_i)^2+ (y-y_i)^2 \ge 1/4, i=1,2,3 \right\} .$$
This flow domain is discretized on an unstructured grid 
with 4225 triangles and 8633 vertices (see Fig.\ \ref{Fig:Pinball:Grid}).
This discretization optimizes the speed of the numerical simulation
while keeping the accuracy at an acceptable level.
Increasing the number of triangles by a factor 4 yields virtually indistinguishable results.
The Navier-Stokes equation is numerically integrated 
with an implicit Finite-Element Method~\citep{Noack2003jfm,Noack2016jfm}.
The numerical integration 
is second-order accurate in space and third-order accurate in time.
This direct numerical simulation has a companion experiment
at turbulent Reynolds numbers $Re_D \approx 4000$--$6000$ \citep{Raibaudo2017ifac}.

\subsubsection{Attractor data}
\label{Sec:Pinball:AttractorData}

\begin{figure}
	\centering
	\includegraphics[height=8cm]{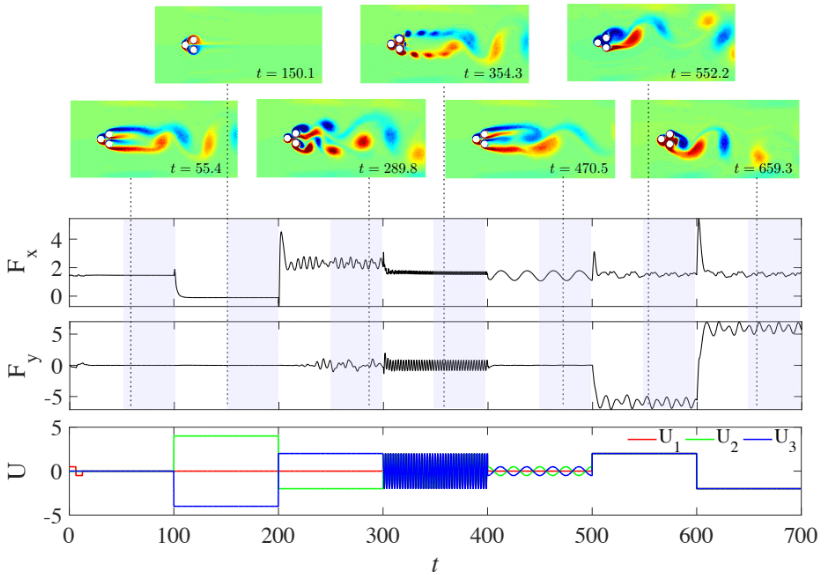}
	\caption{Operating conditions of the fluidic pinball simulation.
                Bottom: time series of the actuation commands for each cylinder over all  phases together.
                Top: the flow of each phase is illustrated 
                with a vorticity snapshot corresponding to maximum lift or minimum drag of the converged phase.
                Center: the time series of the total drag $F_x$ and total lift $F_y$ are displayed.
}
	\label{Fig:Pinball:Simulation}
\end{figure}

We simulate seven actuations in a single simulation
starting at $t=0$ with the unstable steady Navier-Stokes solution.
Each phase is associated with one control law and lasts 100 convective time units.
Figure \ref{Fig:Pinball:Simulation} provides a preview of the simulation
which is detailed in the following. 

In the first phase, vortex shedding is kick-started 
with a motion of the front cylinder,
\begin{subequations}
\label{Eqn:Unforced}
\begin{eqnarray}
U_F &=& \left \{
        \begin{array}{ll}
         1/2  & \quad\quad\quad t \le 6.25 \\ 
        -1/2  & 6.25 < t \le 12.5 \\ 
         0 & \text{otherwise}
        \end{array}
        \right .
\\
U_B &=& -U_T = 0 .
\end{eqnarray}
\end{subequations}
The imposed period $12.5$ corresponds to a Strouhal number of $0.2$ based on the transverse width $5R$.
Without this kick-start, the onset of vortex shedding may require hundreds of shedding periods
depending on accuracy of the steady solution and the truncation errors in the Navier-Stokes solver.
In the  second phase,  
strong boat tailing with symmetric cylinder rotation of the upper and lower cylinder
pushes the separation close to the $x$-axis and completely suppresses vortex shedding,
\begin{equation}
\label{Eqn:BoatTailing}
U_F = 0, \quad U_B = - U_T = 4.
\end{equation}
In the third phase, base-bleed is enforced with opposite cylinder motion:
\begin{equation}
\label{Eqn:BaseBleed}
U_F = 0, \quad U_B = - U_T = -2.
\end{equation}
This actuation widens the wake and pushes vortex formation downstream.
In the fourth phase, 
symmetric high-amplitude, high-frequency actuation energizes both shear layers:
\begin{equation}
\label{Eqn:HighFrequency}
U_F = U_B = U_T = 2 \cos \left(10 \pi t / 12.5 \right).
\end{equation}
The frequency corresponds roughly to five times the one of natural vortex shedding,
following \citet{Thiria2006jfm} for a single cylinder.
In the fifth phase, 
symmetric low-amplitude, low-frequency  forcing delays vortex shedding:
\begin{equation}
\label{Eqn:LowFrequency}
U_F = 0, \quad U_B = - U_T = \frac{1}{2} \cos \left( \pi t / 12.5 \right).
\end{equation}
Low amplitudes were found to be most effective in stabilizing the wake
in a parametric study \citet{Rolland2017limsi}.
In the sixth phase,
a uniform rotation of all cylinders deflects the wake upwards via the Magnus effect,
\begin{equation}
\label{Eqn:MagnusPositive}
U_F =  U_B =  U_T =2.
\end{equation}
In the seventh and last phase,
the opposite Magnus effect is imposed,
\begin{equation}
\label{Eqn:MagnusNegative}
U_F =  U_B =  U_T =-2.
\end{equation}
Table \ref{Tab:Pinball:Simulation} summarizes all control laws for later reference~\footnote{A visualization of the whole simulation can be found on \url{http://fluidicpinball.com}.}.
\begin{table}
	\begin{tabular}{r@{\quad}l@{\quad}l@{\quad}l}
		Phase & Time 
		& Actuation mechanism
		& Control law\\
		\midrule
		I & $t \le 100$ 
		& unforced reference 
		& $U_F=U_B=U_T=0$ for $t>12.5$, see \eqref{Eqn:Unforced} 
		\\    II & $t \in (100,200]$
		& boat tailing
		& $U_F=0, U_B=-U_T=4$
		\\   III & $t \in (200,300]$
		& base bleed
		& $U_F=0, U_B=-U_T=-2$
		\\    IV & $t \in (300,400]$
		& high-frequency forcing
		& $U_F=U_B=U_T=2 \cos (10\pi t / 12.5)$
		\\     V & $t \in (400,500]$
		& low-frequency forcing
		& $U_F=0, U_B=-U_T=1/2 \cos (\pi t / 12.5)$
		\\    VI & $t \in (500,600]$
		& positive Magnus effect
		&  $U_F=U_B=U_T=2$
		\\   VII & $t \in (600,700]$
		& negative Magnus effect
		&  $U_F=U_B=U_T=-2$\\
		\bottomrule         
	\end{tabular}
	\caption{Summary of the different control phases of the fluidic pinball simulation and the associated actuation mechanism.}
	\label{Tab:Pinball:Simulation}
\end{table}

The attractor data 
contains time-resolved snapshots from the last 50 convective units of each phase.
These snapshots are equidistantly sampled with a time step of $0.1$,
i.e., each phase is represented by 500 velocity fields.
The first 50 convective time units correspond 
to $2.5$ downwash times --- enough for the transient dynamics to die out.
We decided to perform one simulation with different control laws
since the transients reveal the robustness of the posttransient phase.

\subsection{Actuated turbulent boundary layer}
\label{Sec:TBL:Plant}

The generation of data of the turbulent boundary layer flow over a surface undergoing a transversal spanwise traveling wave motion is described in this section. 
First, the configuration of the boundary layer flow is described in \S~\ref{Sec:TBL:Configuration}. 
Then, the numerical method of the flow solver is presented in  \S~\ref{Sec:TBL:LES}. 
Finally, the collection of attractor data from 38 simulations 
with varying actuation parameters is detailed in  \S~\ref{Sec:TBL:AttractorData}.

\subsubsection{Boundary layer configuration}
\label{Sec:TBL:Configuration}
The zero-pressure gradient (ZPG) turbulent boundary layer flow actuated by a sinusoidal wall motion is defined in a Cartesian domain with the $x$-axis corresponding to the mean flow direction, the $y$-axis pointing into wall-normal direction, and the $z$-axis in the spanwise direction. Positions are denoted by $\mathbf{x} = (x,y,z)$ and the corresponding velocities by $\mathbf{u} = (u,v,w)$, the pressure is given by $p$ and the density by $\rho$. The governing equations of the flow are the unsteady compressible Navier-Stokes equations and the thermal and caloric state equations. The heat flux is described by Fourier's law and the temperate dependence of the fluid viscosity is given by Sutherland's law. Unlike standard ZPG turbulent boundary layer flow, the actuated flow is statistically three-dimensional due to the wave propagating in the $z$-direction. The flow variables are non-dimensionalized using the flow quantities at rest, the speed of sound $a_0$, and the momentum thickness of the boundary layer at $x_0 = 0$, 
such that $\theta(x_0=0) = 1$. The momentum thickness based Reynolds number is $Re_\theta = 1,000$ at $x_0$ where $Re_\theta = u_{\infty} \theta / \nu$. The Mach number is  $M = 0.1$, i.e., the flow is nearly  incompressible.

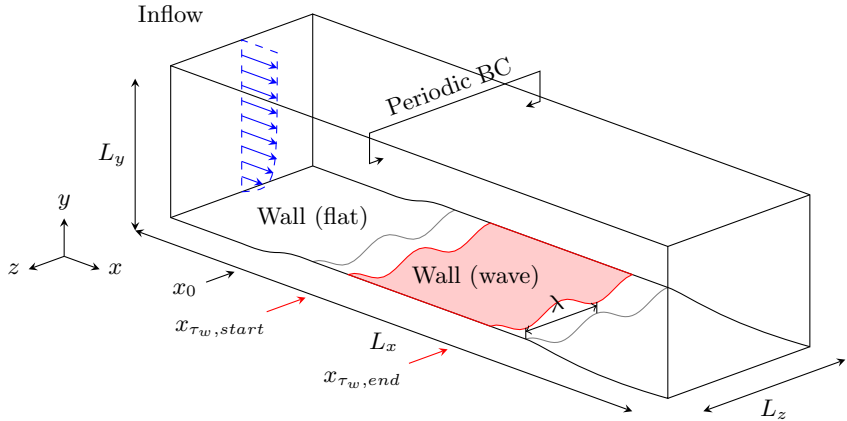
\begin{figure}
  \centering
  \begin{tikzpicture}[x={(0.939cm,-0.34cm)}, y={(0cm,1cm)}, z={(0.939cm,0.34cm)}]\
    \draw [->,>=stealth] (0,0,-1.5) -- (0,0,-2.0) node [left]{$z$};
    \draw [->,>=stealth] (0,0,-1.5) -- (0,0.5,-1.5) node [above]{$y$};
    \draw [->,>=stealth] (0,0,-1.5) -- (0.5,0,-1.5) node [right]{$x$};
    \draw [<->,>=stealth] (0,0,-0.5) -- (7,0,-0.5) node [pos=.5,below=1.0]{$L_x$};
    \draw [<->,>=stealth] (0,0,-0.5) -- (0,2,-0.5) node [pos=.5,left]{$L_y$};
    \draw [<->,>=stealth] (7.5,0,0) -- (7.5,0,2) node [pos=.5,below=2.0]{$L_z$};
    \draw (0,0,0) -- (1,0,0) sin (1.5,0.05,0) cos (2,0.1,0) -- (5,0.1,0) .. controls (5.3,0.1,0) and (5.7,-0.1,0) .. (7,0.0,0) -- (7,2,0) -- (7,2,0) -- (0,2,0) -- cycle;
    \draw (0,0,2) -- (1,0,2) sin (1.5,0.05,2) cos (2,0.1,2) -- (5,0.1,2) .. controls (5.3,0.1,2) and (5.7,-0.1,2) .. (7,0.0,2) -- (7,2,2) -- (7,2,2) -- (0,2,2) -- cycle;
    \draw[color=red,fill=red, fill opacity=0.2, domain=0:2, variable=\z]  (2.5,0.1,0) -- plot (2.5,{0.1*sin((1.5707+2*3.14159*\z) r)},\z)  -- (4.5,0.1,2) --  plot (4.5,{0.1*sin((1.5707+2*3.14159*\z) r)},2.0-\z)  -- (4.5,0.1,0) -- (2.5,0.1,0) -- cycle;
    \draw (7,0,0) -- (7,0,2);
    \draw (0,0,0) -- (0,0,2);
    \draw (0,2,0) -- (0,2,2);
    \draw (7,2,0) -- (7,2,2);
    \draw[opacity=0.5, variable=\z, samples at={0,0.05,...,2.05}]
    plot (5, {0.1*sin((1.5707+2*3.14159*\z) r)}, \z);      
    \draw[opacity=0.5, variable=\z, domain=0:2]
    plot (2, {0.1*sin((1.5707+2*3.14159*\z) r)}, \z);    
    \draw (5,0.1,0) -- (5,0.25,0);
    \draw (5,0.1,1) -- (5,0.25,1);
    \draw[<->] (5,0.2,0) -- (5,0.2,1) node [pos=.5,sloped,above] {$\lambda$};
    
    \node (x0) at (1.5,0,-1.3) {$x_0$};
    \draw[->,>=stealth] (x0) -- (1.5,0,-0.6);
    \node (tau1) at (2.5,0,-1.8) {$x_{\tau_w,start}$};
    \draw[color=red,->,>=stealth] (tau1) -- (2.5,0,-0.6);
    \node (tau2) at (4.5,0,-1.8) {$x_{\tau_w,end}$};
    \draw[color=red,->,>=stealth] (tau2) -- (4.5,0,-0.6);
    \node (inflow) at (-1,2,1) {Inflow};
    \node (flat) at (1, 0,1) {\small{Wall (flat)}};
    \node (flat) at (3.3, 0,1) {\small{Wall (wave)}};
    \draw[<->,>=stealth] (3,1.8,0) -- (3,1.8,-0.2) -- (3,2.2,-0.2) -- node[pos=.5,sloped,above] {Periodic BC} (3,2.2,2.2) -- (3,1.8,2.2) -- (3,1.8,2.0);
    \draw[color=blue, dashed] (0,0,1) .. controls (0.5,0.2,1) ..  (0.5,2,1) -- (0,2,1) -- (0,0,1);
    \draw[color=blue, ->,>=stealth] (0,0.2,1) -- (0.3,0.2,1);
    \draw[color=blue, ->,>=stealth] (0,0.4,1) -- (0.45,0.4,1);
    \draw[color=blue, ->,>=stealth] (0,0.6,1) -- (0.5,0.6,1);
    \draw[color=blue, ->,>=stealth] (0,0.8,1) -- (0.5,0.8,1);
    \draw[color=blue, ->,>=stealth] (0,1.0,1) -- (0.5,1.0,1);
    \draw[color=blue, ->,>=stealth] (0,1.2,1) -- (0.5,1.2,1);
    \draw[color=blue, ->,>=stealth] (0,1.4,1) -- (0.5,1.4,1);
    \draw[color=blue, ->,>=stealth] (0,1.6,1) -- (0.5,1.6,1);
    \draw[color=blue, ->,>=stealth] (0,1.8,1) -- (0.5,1.8,1);
  \end{tikzpicture}
  \caption{Overview of the physical domain of the actuated turbulent boundary layer flow, where $L_x, L_y,$ and $L_z$ are the dimensions of the domain in the Cartesian directions, $\lambda$ is the wavelength of the spanwise traveling wave, $x_0$ marks the onset of the actuation, and $x_{\tau_w,start}$ and $x_{\tau_w,end}$ denote the interval of the integration of the wall-shear stress $\tau_w$.}
  \label{Fig:TBL:Grid}
\end{figure}

An overview of the setup is given in figure~\ref{Fig:TBL:Grid}. The dimensions of the physical domain are $L_x = 190\,\theta$, $L_y = 105\,\theta$, and $L_z = 21.65\,\theta$. At the inflow of the domain, the reformulated synthetic turbulence generation (RSTG) method by \cite{Roidl2013} is used to prescribe a fully turbulent inflow distribution with an adaptation length of less than five boundary-layer thicknesses $\delta_{99}$, such that a natural turbulence state is achieved at $x_0$, which marks the onset of the actuation. Characteristic outflow conditions are applied at the downstream and upper boundary of the domain and periodic conditions are used in the spanwise direction. On the wall, no-slip conditions are imposed and the wall motion is described by
\begin{equation}
  \label{Eq:TBL:Actuation}
y^+|_\mathrm{wall}(z^+,t^+)  = g(x) A^+\cos\left(\frac{2\pi}{\lambda^+} z^+ + \frac{2\pi}{T^+}t^+\right),
\end{equation}
where $A^+ = A u_\tau / \nu$ is the amplitude, $\lambda^+ = \lambda u_\tau / \nu$  the wavelength, and $T^+ = T u_\tau^2 / \nu$  the period of the traveling wave in inner coordinates, i.e., scaled by the kinematic viscosity $\nu$ and the friction velocity $u_\tau(x_0)$ of the non-actuated reference case. The piecewise defined function
\begin{subequations}
\label{Eqn:TBL:Piecewise}
\begin{eqnarray}
g(x) &=& \left \{
  \begin{array}{ll}
    0 & \text{if} \quad x < -5 \\ 
    \frac{1}{2}\left[ 1-\cos\left( \frac{\pi (x+5)}{10}\right) \right] & \text{if} \quad -5 \leq x < 5\\ 
    1  & \text{if} \quad 5 \leq x < 130\\
    \frac{1}{2}\left[ 1+\cos\left( \frac{\pi (x-130)}{10}\right) \right] & \text{if} \quad 130 \leq x < 140\\ 
    0 & \text{otherwise}
  \end{array}
    \right.
\\
\end{eqnarray}
\end{subequations}
enables a smooth streamwise transition from a flat non-actuated to an actuated wall and vice versa. Apart from the reference case without any wall actuation, i.e., $A^+ = 0$ and $T^+ = 0$, 37 parameter combinations of $\lambda^+$, $T^+$, and $A^+$ are considered (see table~\ref{Tab:TBL:Simulation}). Most parameter points were generated using latin hypercube sampling.

The physical domain is discretized by a structured body-fitted grid with a resolution of $\Delta x^+ = 12$ in the $x$-direction, $\left. \Delta y^+ \right|_{wall} = 1.0$ in the $y$-direction using gradual coarsening with increasing distance from the wall, and $\Delta z^+ = 4.0$ in the $z$-direction. This nearly DNS-like resolution guarantees to capture all relevant turbulent scales and allows a smooth representation of the wavy wall. In total, the grid consists of $n = 732 \times 131 \times 250 \approx 24 \cdot 10^6$ cells. The details of the flow conditions and grid parameters are summarized in table~\ref{Tab:TBL:GridParams}.

\begin{table}
  \centering
  \begin{tabularx}{0.6\textwidth}{l@{\quad}l@{\quad}}
    Parameter & Value \\
    \midrule
    $M_{\infty}$ & $0.1$ \\
    $Re_{\theta}$ & $1,000$\\
    $\Delta x^+$ & $12.0$\\
    $\Delta y^+_{wall}$ & $1.0$\\
    $\Delta z^+$ & $4.0$\\
    $\delta_{99} (x=0.0))$ & $8.638$\\
    $\delta_1 (x=0.0)$ & $1.488$\\
    $\theta (x=0.0)$ & $1.0$\\
    $L_x \times L_y \times L_z$ & $190.15 \times 104.91 \times 21.65$\\
    $n_{\text{cells}}$ & $732 \times 131 \times 250 \approx 24 \cdot 10^6$\\
    $x_{\tau_w,start}$ & $50.0$\\
    $x_{\tau_w,end}$ & $100.0$\\
    \bottomrule
  \end{tabularx}
  \caption{Flow and grid parameters of the non-actuated reference case and the 37 actuated cases.}
  \label{Tab:TBL:GridParams}
\end{table}

\subsubsection{Large-eddy simulation solver}
\label{Sec:TBL:LES}
The actuated turbulent boundary layer flow is simulated using a finite volume approximation of the unsteady compressible Navier-Stokes equation on a structured body-fitted mesh. A second-order accurate formulation of the inviscid fluxes using the advection upstream splitting method (AUSM) by~\cite{Liou1993} is applied. The cell-surface values of the flow quantities are reconstructed from the surrounding cell-center values using a Monotone Upstream Scheme for Conservation Laws (MUSCL) type strategy. The viscous fluxes are discretized by a modified cell-vertex scheme at second-order accuracy. The time integration is performed by a second-order accurate five-stage Runge-Kutta scheme, rendering the overall discretization second-order accurate. The subgrid scales in the LES are implicitly modeled following the  monotonically integrated large-eddy simulation approach~\citep{Boris1992}, i.e., the numerical dissipation of the AUSM scheme models for the viscous dissipation of the high wavenumber turbulence spectrum \citep{Meinke2002}. Thus, the small-scale structures are not explicitly resolved and the grid is used as a spatial filter resolving the large energy-containing structures in the inertial subrange. To capture the temporal variation of the geometry, the Navier-Stokes equations are written in the Arbitrary Lagrangian-Eulerian (ALE) formulation \citep{Hirt1997} such that the actuated wall can be represented by an appropriate mesh deformation. Additional volume fluxes are determined to satisfy the Geometry Conservation Law (GCL).

The numerical method has been  thoroughly validated by computing a wide variety of internal and external flow problems~\citep{Ruetten2005,Alkishriwi2006,Renze2008,Statnikov2017}. Analyses of drag reduction in turbulent boundary layer flow have been performed for riblet structured surfaces~\citep{Klumpp2010a} and for traveling transversal surface waves~\citep{Klumpp2011,Koh2015a,Koh2015,Meysonnat2016}. The quality of the results confirms the validity of the approach for the current flow problem.

\subsubsection{Attractor data}
\label{Sec:TBL:AttractorData}
\begin{figure}
  \centering
\includegraphics[height=7cm]{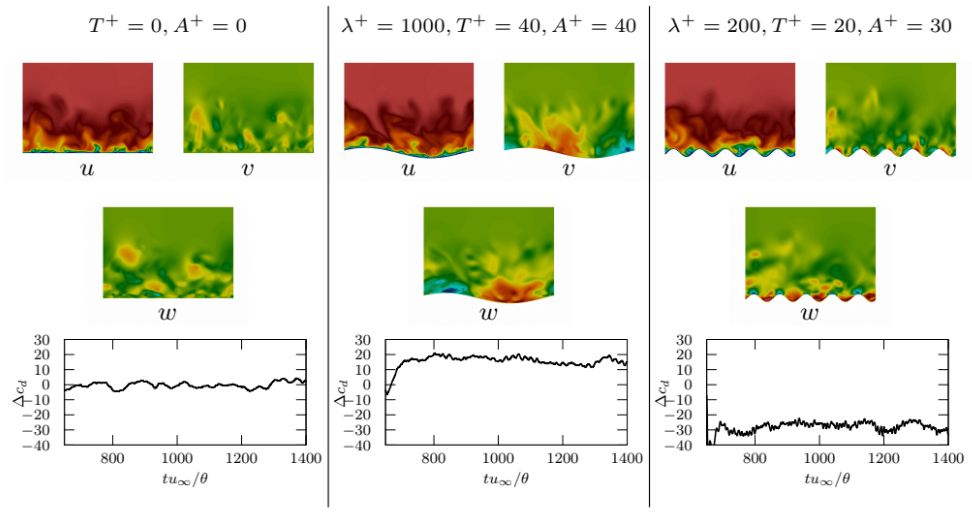}
\caption{Illustration of the turbulent boundary layer flow: (left) non-actuated reference case $N_1$, (center) actuated case with highest drag reduction $N_{24}$, and (right) actuated case with lowest drag reduction $N_2$;
  (top) contour plots of the instantaneous Cartesian velocity components $u$,$v$, and $w$ in a $y-z$ plane at $x/\theta \approx 65$;
  (bottom) time evolution of the instantaneous relative drag reduction $\Delta c_d$.
}
\label{Fig:TBL:Simulation}
\end{figure}
\begin{figure}
	\centering
  \includegraphics[height=8cm]{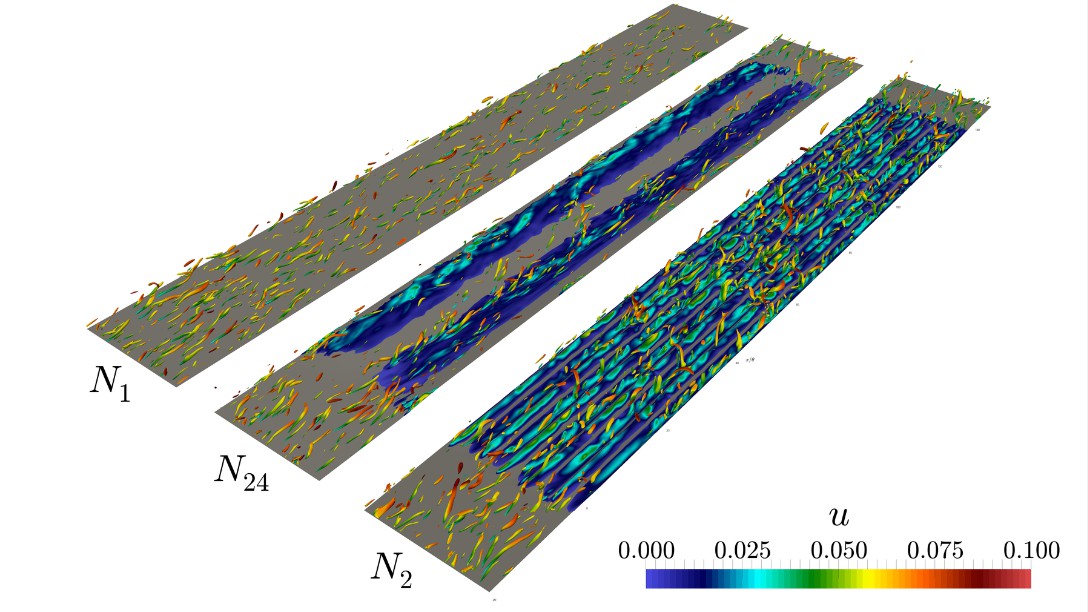}
	\caption{Contour plot of the $\lambda_2$-criterion \citep{Jeong1995jfm}, coloured by the instantaneous streamwise velocity, for three turbulent boundary layer flows; non-actuated reference case $N_1$, actuated highest drag reduction case $N_{24}$, actuated lowest drag reduction case $N_2$.}
\label{Fig:TBL:Lambda2}
\end{figure}

First, the simulation of the reference setup is run for $t u_\infty / \theta \approx 650$ convective units until a quasi-steady state of the drag evolution is observed. 
Then, the average drag of the reference setup is measured for the next $\Delta t u_\infty / \theta \approx 1000$ convective units. Subsequently, the actuated cases are initiated using an intermediate solution from the reference case and a quick transition from the flat wall to the fully deflected wall is enforced. When a new quasi-steady state of the friction drag is obtained, i.e., after $\Delta t u_\infty / \theta \approx 150$ convective units,  the drag of each actuated case is averaged over a period of $\Delta t u_\infty / \theta \approx 800$ convective units. 
The relative drag reduction for the i-th test case
\begin{equation}
  \label{eq:1}
\Delta c_{d,\mathrm{i}} = \frac{\int\limits_{A,\mathrm{na}} \tau_{w,\mathrm{na}} dA - \int\limits_{A,\mathrm{i}} \tau_{w,\mathrm{i}} dA}{\int\limits_{A,\mathrm{na}} \tau_{w,\mathrm{na}} dA} \cdot 100
\end{equation}
is computed by integrating the wall-shear stress distribution $\tau_w$ over the wetted surface $A$ in the interval $x \in [x_{\tau_w,start}, x_{\tau_w,end}]$ of the non-actuated reference (subscript $na$) and the actuated setups (subscript $i$). The values of the relative drag reduction $\Delta c_d$ and the skin-friction reduction $\Delta c_f$, i.e., the drag reduction without considering the increase in the wetted surface, of the various amplitude, wavenumber, and period prescriptions are listed in table~\ref{Tab:TBL:Simulation}. An exemplary illustration of the impact of the moving surface on the near-wall velocity field and the friction drag development is given in figure~\ref{Fig:TBL:Simulation} for the reference case $N_1$, the case with the highest drag reduction $N_{24}$, and the case with the lowest drag reduction $N_2$. In figure~\ref{Fig:TBL:Lambda2}, the variation of the turbulent structures in the near-wall region of these three cases is shown. The comparative juxtaposition of the data evidences the decrease of the turbulent structures for the $N_{24}$ case.
\begin{table}
  \centering
  \begin{tabularx}{0.55\textwidth}{r@{\qquad}r@{\qquad}r@{\qquad}r@{\qquad}r@{\qquad}r}
    $N$ & $\lambda^+$ & $T^+$ & $A^+$ & $\Delta c_d$ & $\Delta c_f$ \\
    \midrule
    1  & -- & -- & -- & -- & -- \\
    & & & & & \\
    2  & 200 & 20 & 30 & -27.57 & -6.82 \\
    3  & 200 & 30 & 21 & -1.13 & 8.16 \\
    4  & 200 & 40 & 30 & -9.43 & 8.37 \\
    5  & 200 & 50 & 45 & -26.58 & 8.93 \\
    6  & 200 & 60 & 30 & -9.89 & 7.98 \\
    7  & 200 & 70 & 14 & -0.79 & 3.70 \\
    8  & 200 & 70 & 38 & -17.42 & 9.23 \\
    9  & 200 & 100 & 28 & -9.79 & 6.29 \\
    & & & & & \\
    10  & 500 & 20 & 30 & -0.17 & 3.18 \\
    11  & 500 & 30 & 22 & 8.18 & 9.88 \\
    12  & 500 & 40 & 21 & 6.41 & 7.99 \\
    13  & 500 & 40 & 30 & 7.51 & 10.61 \\
    14  & 500 & 60 & 30 & 3.77 & 7.00 \\
    15  & 500 & 70 & 36 & 2.66 & 7.24 \\
    16  & 500 & 70 & 64 & -10.52 & 3.56 \\
    17  & 500 & 100 & 48 & -4.53 & 3.70 \\
    & & & & & \\
    18  & 1000 & 20 & 10 & 3.58 & 3.67 \\
    19  & 1000 & 20 & 30 & 11.95 & 12.72 \\
    20  & 1000 & 20 & 50 & -0.45 & 1.93 \\
    21  & 1000 & 40 & 10 & 3.15 & 3.24 \\
    22  & 1000 & 40 & 20 & 6.48 & 6.84 \\
    23  & 1000 & 40 & 30 & 11.79 & 12.56 \\
    24  & 1000 & 40 & 40 & 15.69 & 16.98 \\
    25  & 1000 & 40 & 50 & 14.77 & 16.79 \\
    26  & 1000 & 40 & 60 & 12.49 & 15.42 \\
    27  & 1000 & 80 & 10 & 0.65 & 0.75 \\
    28  & 1000 & 80 & 20 & 3.49 & 3.87 \\
    29  & 1000 & 80 & 30 & 5.60 & 6.42 \\
    30  & 1000 & 80 & 40 & 9.18 & 10.58 \\
    31  & 1000 & 80 & 50 & 8.86 & 11.01 \\
    32  & 1000 & 80 & 60 & 8.34 & 11.41 \\
    33  & 1000 & 120 & 10 & 0.73 & 0.83 \\
    34  & 1000 & 120 & 20 & -0.51 & -0.11 \\
    35  & 1000 & 120 & 30 & 2.07 & 2.93 \\
    36  & 1000 & 120 & 40 & 4.34 & 5.81 \\
    37  & 1000 & 120 & 50 & 3.02 & 5.31 \\
    38  & 1000 & 120 & 60 & 2.03 & 5.31 \\
    \bottomrule
  \end{tabularx}
\caption{Actuation parameters of the turbulent boundary layer simulations, where each setup is denoted by a case number $N$. 
The quantity $\lambda^+$ is the spanwise wavelength of the traveling wave, $T^+$ is the period, and $A^+$ is the amplitude, all given in inner units, i.e., non-dimensionalized with the kinematic viscosity $\nu$ and the friction velocity $u_\tau$. Each block includes setups with varying period and amplitude for a constant wavelength. The list includes the values of the averaged relative drag reduction $\Delta c_d$, and the averaged relative skin friction reduction  $\Delta c_f$.}
\label{Tab:TBL:Simulation}
\end{table}

The snapshots for the computation of the POD modes are obtained in a subdomain spanning from $x = 15 \theta$ to $x = 120 \theta$, from the wall to a thickness of $y = 15.4 \theta$, and over the full spanwise extent of the computational domain. That is, the subdomain covers the complete boundary layer over the actuated part of the wall, excluding the zones of spatial transition. The data are collected at the quasi-steady states with a sampling period of $\Delta t u_\infty / \theta \approx 0.94$, i.e., every $n = 300$ iteration steps.
\section{Comparison methodology for different attractors}
\label{Sec:Method}

In this section, 
we propose a comparison methodology for attractor data.
The first constitutive element is a standard metric 
for snapshots as described in \S~\ref{Sec:Method:Distance}.
In \S~\ref{Sec:Method:Metric},
this metric is generalized to attractor data
and referred to as \emph{Metric for Attractor Overlap} (MAO).
In \S~\ref{Sec:Method:Proximity},
the closeness of all attractors are featured in proximity maps.

A coarse-grained version of MAO
is enabled by clustering (\S~\ref{Sec:Method:Clustering}).
This coarse-graining (\S~\ref{Sec:Method:Overlap})
reduces the computational expense of the metric
and gives visual access to select coherent structures
which the attractors have in common, i.e., they overlap, 
or do not share, i.e., they are disjoint. 

\begin{figure}
	\centering
	\includegraphics[width=140mm]{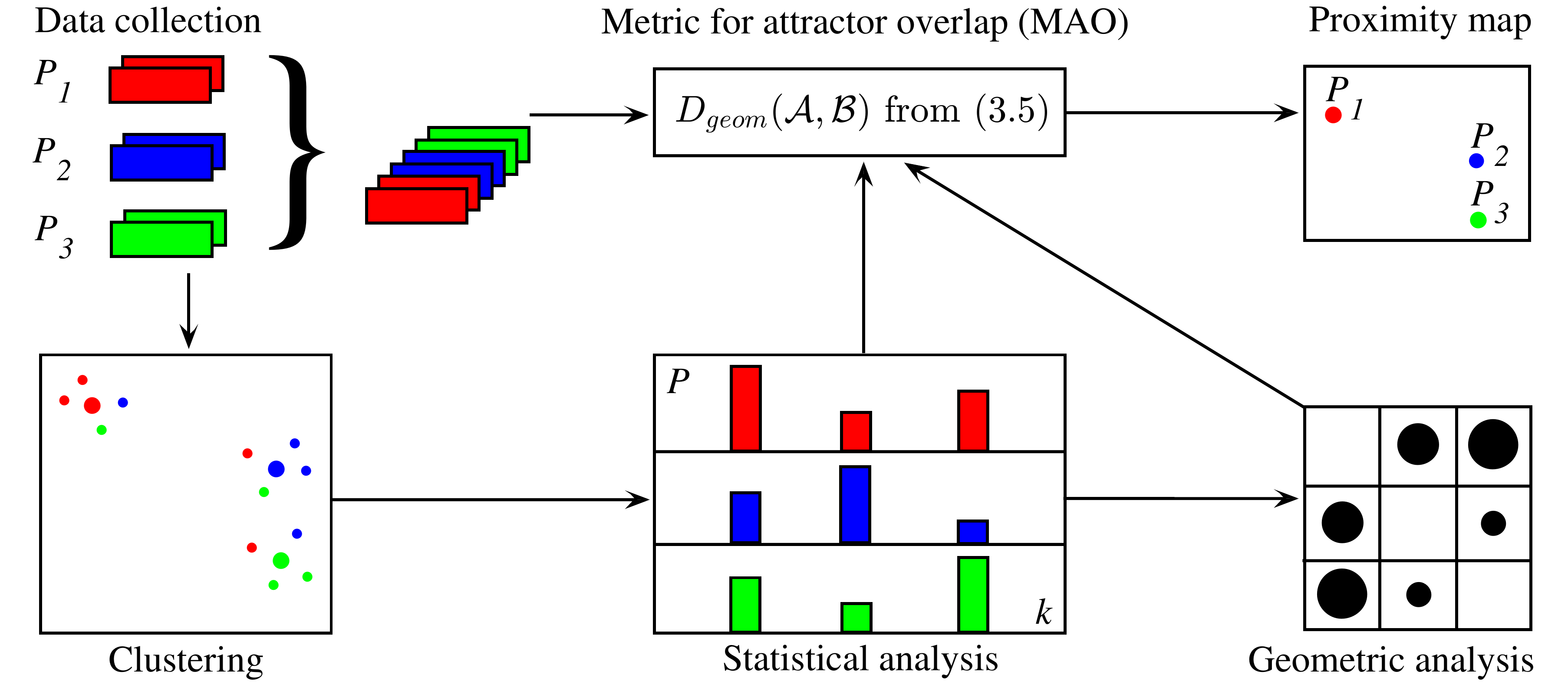}\\ 
	\includegraphics[width=130mm]{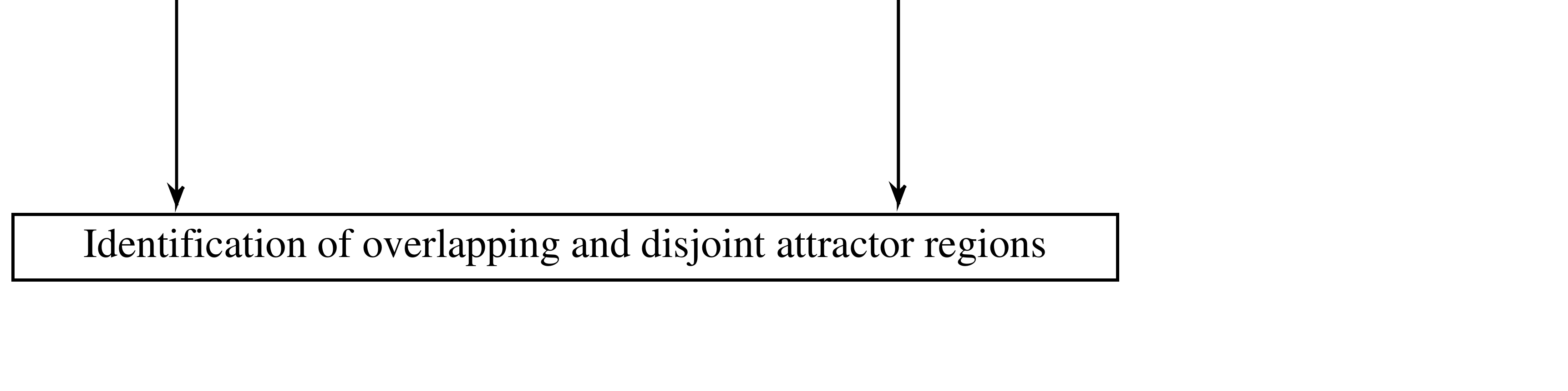}\\ 
	\caption{Schematic of the pipeline for the metric of attractor overlap and associated proximity maps. 
Here, $P_1$, $P_2$ and $P_3$ refer to  data sets outlined in \S~\ref{Sec:Plant}, 
i.e., sets of snapshots which are statistically representative 
for an attractor under a given operating condition. 
Of course, the number of data sets is typically much larger.
$D$ is the metric of attractor overlap (\S~\ref{Sec:Method:Metric}).
Its two arguments $P$, $Q$ represent data sets, like $P_1$, $P_2$ and $P_3$.
The proximity map (top right) is explained in \S~\ref{Sec:Method:Proximity}.
The coarse-graining by clusters (center row) is discussed in \S~\ref{Sec:Method:Clustering}
and the identification of overlapping and disjoint attractor regions (bottom row) 
is outlined in \S~\ref{Sec:Method:Overlap}.}
\label{Fig:Method}
\end{figure}

Figure \ref{Fig:Method} previews the proposed methodology.
The top row illustrates the processing 
from the raw data of multiple attractors (left) to the proximity map (right)
employing the snapshot MAO. 
The center row represents the coarse-grained version with clustering.
Clustering opens the opportunity to pinpoint the overlap region of attractors 
to a few select velocity fields, i.e., shared centroids, as indicated in the bottom row.

\subsection{Distance between snapshots}
\label{Sec:Method:Distance}
Let $\vec{u}(\vec{x})$ and $\vec{v}(\vec{x})$ be two  velocity fields
in the domain $\Omega$.
We define the distance between these fields as 
\begin{equation}
\label{Eqn:DistanceSnapshots}
D(\vec{u},\vec{v}) = \sqrt{\int\limits_{\Omega}\!\! d\vec{x} \> \Vert \vec{u}(\vec{x}) -\vec{v} (\vec{x}) \Vert^2} .
\end{equation}
Here, $\Vert \cdot \Vert$ denotes the Euclidean norm.
Note that this distance is based on the norm associated with the Hilbert space 
$\mathcal{L}^2  ( \Omega)$ of square-integrable functions.
It fulfills all properties of a metric, 
like positive definiteness, 
commutativity, 
scaling with a real factor 
and triangle inequality.
For the fluidic pinball data of \S~\ref{Sec:Pinball:Plant}, 
the distance measure uses the whole computational domain $\Omega$.
For the turbulent boundary layer of \S~\ref{Sec:TBL:Plant}, 
a subdomain over the whole actuated surface is chosen.
The surface actuation leads to a small domain deformation.
This deformation is neglected and the operations are performed
in a stationary domain in which the grid points assume  
their the unactuated equilibrium location.

\subsection{Metric of attractor overlap}
\label{Sec:Method:Metric}

In this section, we define a measure for the attractor similarity
based on the snapshot configuration and the Hilbert-space metric.
Loosely speaking, the difference between two attractors $\mathcal{A}$ and $\mathcal{B}$
--- represented by their snapshot ensembles --- 
is geometrically defined as the sum of the average distances 
between the snapshots of $\mathcal{A}$ to $\mathcal{B}$ and vice versa.

In the following, this quantity is defined.
For simplicity, let us consider 
two attractors $\mathcal{A}$ and $\mathcal{B}$.
Let $\mathcal{M} := \{ \bm{u}^m \}_{m=1}^M$ be the union of all snapshots from both attractors.
For simplicity, we assume the generic case that all snapshots are pairwise different.
The subset of $\mathcal{M}$ belonging to attractor $\mathcal{A}$
is defined by the characteristic function
\begin{equation}
\label{Eqn:CharacteristicFunctionAttractor}
\chi^m_\mathcal{A}  :=  \begin{cases}
             1, &  \mbox{if } \vec{u}^m \in \mathcal{A},\\
             0, &  \mbox{otherwise}.
             \end{cases}
\end{equation}
The number of snapshots in $\mathcal{A}$ is given by
$M_A := \sum_{m=1}^M \chi^m_\mathcal{A} $.
Similar formulae hold for attractor $\mathcal{B}$.

The distance of a snapshot $\vec{u}$ to  $\mathcal{B}$
is defined by the closest corresponding snapshot of $\mathcal{B}$:
\begin{equation}
D(\vec{u},\mathcal{B}) = \min\limits_{m=1,\ldots,M \atop \chi^m_\mathcal{B}=1}  D(\vec{u}, \vec{u}^m ).
\end{equation}
The average distance of attractor $\mathcal{A}$ to $\mathcal{B}$ is defined by 
\begin{equation}
D(\mathcal{A},\mathcal{B}) = \frac{1}{M_A} 
\sum\limits_{m=1,\ldots,M \atop \chi^m_\mathcal{A}=1}  D\left( \vec{u}^m, \mathcal{B} \right).
\end{equation}
This distance is not commutative.
Suppose $\mathcal{A}$ has only one snapshot of $\mathcal{B}$ and $\mathcal{B}$ has many more elements.
In this case,  $D(\mathcal{A},\mathcal{B})=0$ but $D(\mathcal{B},\mathcal{A})>0$.
Hence, we define a symmetrized version of this distance
\begin{equation}
\label{Eqn:MAO:Snapshots}
D_{\rm geom} (\mathcal{A},\mathcal{B} ) :=  \frac{D \left(\mathcal{A},\mathcal{B} \right) 
                                                + D \left(\mathcal{B},\mathcal{A} \right)}{2}.
\end{equation}
We refer to this quantity as \emph{Metric of Attractor Overlap (MAO)}.
MAO has the properties of 
a \emph{metric} for snapshot ensembles $\mathcal{A}$ and $\mathcal{B}$ and $\mathcal{C}$
since the defining  properties can be shown for the generic case of pairwise different snapshots.
\begin{description}
\item[(1) Positive definiteness:]\ $D_{\rm geom}(\mathcal{A},\mathcal{B}) \ge 0$ for all $\mathcal{A},\mathcal{B}$
and 
$$\mathcal{A} = \mathcal{B} \quad \Leftrightarrow \quad D_{geom} \left(\mathcal{A},\mathcal{B} \right) =0.$$ 
\item[(2) Symmetry:]\  By definition \eqref{Eqn:MAO:Snapshots},
$$D_{\rm geom} \left(\mathcal{A}, \mathcal{B} \right) = D_{\rm geom} \left( \mathcal{B}, \mathcal{A} \right).$$
\item[(3) Triangle inequality:] 
$$D_{\rm geom} \left(\mathcal{A}, \mathcal{C} \right) \le 
  D_{\rm geom} \left(\mathcal{A}, \mathcal{B} \right) + 
  D_{\rm geom} \left(\mathcal{B}, \mathcal{C} \right).$$
\end{description}

To illustrate the physical implications of MAO, we consider several cases.
If $\mathcal{A}$ and $\mathcal{B}$ only consist
of velocity fields $\vec{u}$ and $\vec{v}$, respectively,
their distance coincides with the Hilbert-space metric,
$D(\mathcal{A}, \mathcal{B}) = D(\vec{u},\vec{v})$

Let $\mathcal{A}$ contain one velocity field, 
e.g., an unstable steady solution, 
and $\mathcal{B}$ contains many fields, e.g., snapshots of a stable periodic dynamics.
Then  $D(\mathcal{A},\mathcal{B})$ is the \emph{smallest} distance between the steady solution and the limit cycle,
while $D(\mathcal{B},\mathcal{A})$ represents the \emph{average} distance between fixed point and limit cycle.
This asymmetry makes sense. 
$D(\mathcal{A},\mathcal{B})$ quantifies how well elements of $\mathcal{A}$ 
can be represented by elements of $\mathcal{B}$ \emph{on average}.
This measure is inherently non-communitative:
A small set may be better represented by a rich set than the other way round.
MAO  is the average of both,
i.e.,  an effective average distance between two attractors,
not the minimal geometric distance between the closest elements of two sets.

Next, 
let us assume that both attractors arise from periodic dynamics 
defining limit cycles with the same origin  and the same plane
but with radii $10$ and $11$.  
In this case, the distance approaches unity for sufficiently large amount of snapshots.
Complete attractor overlap, 
$D_{\rm geom} \left( \mathcal{A}, \mathcal{B} \right) = 0$, 
implies that $\mathcal{A}$ and $\mathcal{B}$ have the same snapshots.
The probability of their occurrence may differ in both data.

In summary, MAO  averages how well each snapshot of $\mathcal{A}$
is represented by the closest snapshot of $\mathcal{B}$ and vice versa.
Note that some of the above statements need to be refined in case of identical snapshots.
We shall not pause to do so.

\subsection{Proximity map}
\label{Sec:Method:Proximity}
In case of two attractors $\mathcal{A}$ and $\mathcal{B}$,
the closeness can be characterized by MAO, i.e., a single number.
The comparison of many attractors $\mathcal{A}^l$, $l=1,\ldots,L$ with  $L \gg 1$ is more challenging.
In this case,  the complete snapshot set $\left\{ \vec{u}^m \right\}_{m=1}^M$
comprises the elements of all $L$ attractors.
The closeness of the attractors $\mathcal{A}^l$ 
may be visualized in a two-dimensional proximity map
which preserves the metric of attractor overlap as good as possible
(see figure \ref{Fig:Method} \textsl{top, right}).
This task is performed  by multi-dimensional scaling  \citep{Cox2000book}
presented in the following.

The relative distances between the attractors is expressed by the symmetric matrix
\begin{equation}
\label{Eqn:AttractorDistanceMatrix}
D_{ln} := D_{\rm geom} \left( \mathcal{A}^l, \mathcal{A}^n \right), \quad l,n=1, \ldots, L.
\end{equation}
Let $ \vec{\gamma}^l \in \mathbb{R}^2$ 
be a two-dimensional feature vector associated with the $l$th operating condition.
The goal of a proximity map 
is to find a mapping $\mathcal{A} \mapsto \vec{\gamma}$
such that the pointwise distances in the feature plane are preserved as good as possible,
\begin{equation}
\sum\limits_{l,n=1}^L  \left(  D_{ln} - \left \Vert \vec{\gamma}^l-\vec{\gamma}^n \right \Vert \right)^2 
\overset{!}{=} \hbox{min}.
\end{equation}
At this point, the feature plane is indeterminate with respect to translation, rotation, and reflection.
However, one can request that the center of the feature vector is at the origin,
removing the translative degree of freedom.
Moreover, the variances of the first feature coordinate can be maximized,
removing the rotational degree of freedom.
Now, 
only the indeterminacy with respect to reflection is left, 
like for POD modes.

The proximity map can easily be generalized for higher-dimensional feature spaces.
Moreover, the proximity map requires only a distance matrix 
and can thus also be used for snapshots  using the snapshot metric \eqref{Eqn:DistanceSnapshots}.
Proximity maps have been presented
for mixing layer and Ahmed body wake data \citep{Kaiser2014jfm} and
for ensembles of control laws \citep{Kaiser2017tcfd,Duriez2016book,Kaiser2017ifac}.

\subsection{Cluster analysis}
\label{Sec:Method:Clustering}

Next, 
the $M$ snapshots $\vec{u}^m (\vec{x})$, $m=1,\ldots,M$, are coarse-grained
into $K$ representative centroids $\vec{c}_k (\vec{x})$, $k=1,\ldots,K$.
These centroids are chosen 
to minimize the total variance
of the snapshots $\vec{u}^m$ with respect to the nearest centroid $\vec{c}_k$,
\begin{equation}
\label{Eqn:MinimumVariation}
V = \sum\limits_{k=1}^K \sum\limits_{\vec{u}^m\in \mathcal{C}_k} D^2 \left(\vec{u}^m, \vec{c}_k \right)
\overset{!}{=} \hbox{\rm min}.
\end{equation} 
Each centroid $\vec{c}_k$ defines a cluster $\mathcal{C}_k$ 
containing all flow states $\vec{u}^m$ 
which are closer to $\vec{c}_k$
as compared to any other centroid $\vec{c}_j$, $j\not = k$.  
Thus, each snapshot $\vec{u}^m$ can be attributed to one cluster $\mathcal{C}_k$.
This cluster affiliation is coded as characteristic function
\begin{equation}
 \label{Eqn:CharacteristicFunction}
T_{k}^{m} := \begin{cases}
                1, &  \mbox{if } \vec{u}^m \in \mathcal{C}_k,\\
                0, &  \mbox{otherwise}.
               \end{cases}
\end{equation}
The number of snapshots in the $k$th cluster reads
\begin{equation}
 N_k = \sum\limits_{m=1}^M\,T_{k}^{m}.
\end{equation}

The centroids can also be expressed in terms of this characteristic function.
It can be shown that they are the mean of all snapshots in the corresponding cluster,
\begin{equation}
\label{Eqn:Centroid}
\vec{c}_k
= \frac{1}{N_k} \sum\limits_{\vec{u}^m \in \mathcal{C}_k}\, \vec{u}^m
= \frac{1}{N_k} \sum\limits_{m=1}^M  \> T_{k}^{m} \vec{u}^m.
\end{equation}

Numerically, the optimization problem \eqref{Eqn:MinimumVariation} for the centroids
is solved using k-means clustering~\citep{Steinhaus1956,MacQueen1967proc,Loyd1982ieee}
and k-means++ \citep{Arthur2007proc} for the initialization. 
Since k-means shows a dependence on the initial conditions, 
we run the corresponding MATLAB routine for $1000$ initial conditions and select the one having the smallest variance.
The iteration stops when convergence is reached,
i.e., the characteristic function \eqref{Eqn:CharacteristicFunction} does not change.
The number of iterations is limited to $10,000$ iterations.

For practical reasons, we perform a lossless POD decomposition prior to the clustering of $M$ snapshots 
into the mean flow $\vec{u}_0$ and $M-1$ modes $\vec{u}_i$, $i=1,\ldots,M-1$
(see \S \ref{Sec:POD}).
Let $\vec{u} = \vec{u}_0 + \sum_{m=1}^{M-1} a_i \vec{u}_i$ be one snapshot
and  $\vec{v} = \vec{u}_0 + \sum_{m=1}^{M-1} b_i \vec{u}_i$ another one.
Then, $D^2 (\vec{u},\vec{v}) = \sum_{m=1}^{M-1} (a_i - b_i)^2$.
The right-hand side requires $3 M$ floating point operations
while the computational load of the integral \eqref{Eqn:DistanceSnapshots} scales with the number of grid points
and is much larger even for the employed grids.

The clustering enables to characterize each operating condition by a corresponding probability distribution.
Let $l=1, \ldots, L$ be the index of the $L$ operating conditions.
Let $N_k^l$ be the number of snapshots of the $l$th attractor data in the $k$th cluster.
Let $N^l = \sum_{k=1}^K N_k^l$ be the total number of snapshots of the $l$th attractor data.
Then, the probability that a snapshot belonging to the $l$th operating condition
lies in the $k$th cluster reads
\begin{equation}
 \label{Eqn:Frequency}
 P_k^l = \frac{N_k^l}{N^l}, \quad k=1,\ldots,K.
\end{equation}
In particular, the probability distribution is determined here as relative frequencies of cluster visits
and fulfills the non-negativity, $P_k^l \ge 0$, and normalization condition, $\sum_{k=1}^K P_k^l = 1$.

\subsection{Cluster-based analysis of the attractor overlap and dissimilarity}
\label{Sec:Method:Overlap}
In this section, we define a measure for the attractor similarity
based on the centroid configuration.
Each snapshot $\vec{u}^m$ is represented by its closest centroid $\vec{c}_k$.
Let $\mathcal{A}$ and $\mathcal{B}$ be two sets of  attractor data
and $\vec{P}$ and $\vec{Q}$ be the corresponding cluster-based probability distributions.
The snapshot-based metric
is coarse-grained to the  average distance of the centroids of attractor $\mathcal{A}$ and $\mathcal{B}$:
\begin{equation}
\label{Eqn:MAO:Centroids}
D_{\rm geom} (\vec{P},\vec{Q}) 
= \frac{1}{2} \sum\limits_{k=1}^K \left[ P_k \>  D(\vec{c}_k,\mathcal{B}^{c})  + Q_k \>  D(\vec{c}_k,\mathcal{A}^{c})  \right] .
\end{equation}
Here $\mathcal{A}^c$ and $\mathcal{B}^c$ denote the centroid ensembles 
associated with attractors $\mathcal{A}$ and $\mathcal{B}$, respectively. 
This formula can be derived from \eqref{Eqn:MAO:Snapshots}
be replacing the snapshots $\bm{u}^m$ by the closest centroids $\vec{c}_k$
and taking into account the population size.
The center row of figure \ref{Fig:Method} 
illustrates the approximative cluster-based analysis.
Evidently, \eqref{Eqn:MAO:Centroids} is an approximation of \eqref{Eqn:MAO:Snapshots}
and is equivalent 
for the  maximum cluster resolution $K=M$  of snapshots.
In this case, each snapshot represents a centroid
and the probability distributions associated with each cluster are unit vectors.

The main computational load for MAO comes from the distance matrix
of the snapshots or centroids.
The computational savings of the cluster-based measure \eqref{Eqn:MAO:Centroids} 
with respect to snapshot-based metric \eqref{Eqn:MAO:Snapshots} scale with $(K/M)^2$.
For the fluidic pinball study with  $M=3500$ snapshots and $K=50$ clusters, 
this translates to the quite dramatic factor of $\approx 2/10000$.
This saving does not include the operations for clustering.
These can be large for a  field-based clustering 
of \S~\ref{Sec:Method:Clustering}
and small for an alternative clustering, 
e.g., based on few aerodynamic force components.

A lossless POD reduces the computational load of the distance matrix
and of the clustering to a tiny fraction of the original cost.
The main cost of POD originates from the correlation matrix 
which requires a similar amount of operations as the distance matrix.
The advantage of POD preprocessing 
is the many  post-processing options at negligible cost.

As a side note, a commonly used metric for probability distributions
is the Jensen-Shannon distance (see appendix \S~\ref{Sec:JensenShannon}).
This distance measures probability differences on shared clusters 
but is blind to the geometric distances of disjoint clusters.
In the previous example of concentric co-planar circles,
the metric is the same for all non-identical radii.
\section{Comparison of fluidic pinball attractors with different control laws}
\label{Sec:Pinball}

The comparison of fluidic pinball attractors is performed
following the proposed methodology of the previous section.
For simplicity and consistency, we follow the coarse-grained data comparison
displayed in the center and bottom row of figure \ref{Fig:Method}.
Firstly (\S~\ref{Sec:Pinball:Clustering},
the cluster analysis is performed.
Sections \S~\ref{Sec:Pinball:Metric} and \S~\ref{Sec:Pinball:Proximity}
present the metric of attractor overlap and associated proximity maps.
It should be noted that the analysis 
is only based on converged post-transient data
while the shown temporal dynamics includes also the actuation transients.

\subsection{Cluster analysis}
\label{Sec:Pinball:Clustering}
\begin{figure}
	\centering
        \includegraphics[width=110mm]{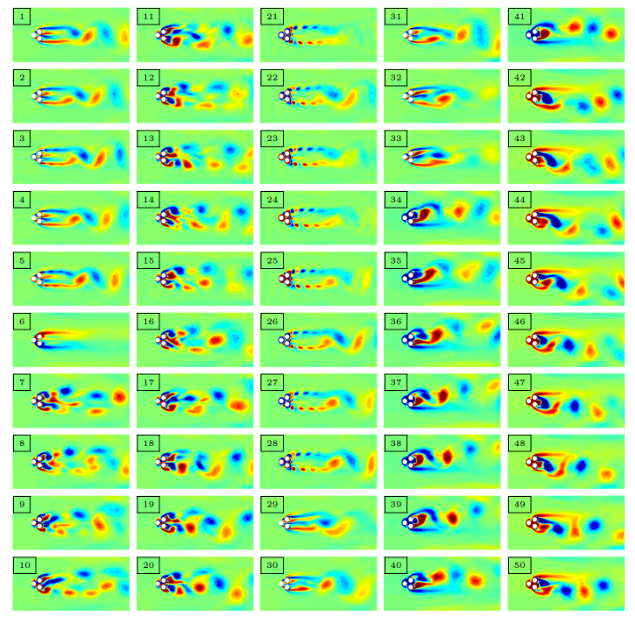}
	\caption{Cluster centroids of the fluidic pinball simulation.
        The centroids are sorted in order of appearance 
        neglecting the first 50 time units (transients) of each phase.
        The visualization depicts the vorticity distribution:
        green, red and blue represent vanishing, positive and negative values, respectively.}
\label{Fig:Pinball:Centroids}
\end{figure}
Figure \ref{Fig:Pinball:Centroids} illustrates $50$ centroids
distilled from $7 \times 500$ post-transient snapshots. 
\begin{figure}
	\centering
	\includegraphics[width=120mm]{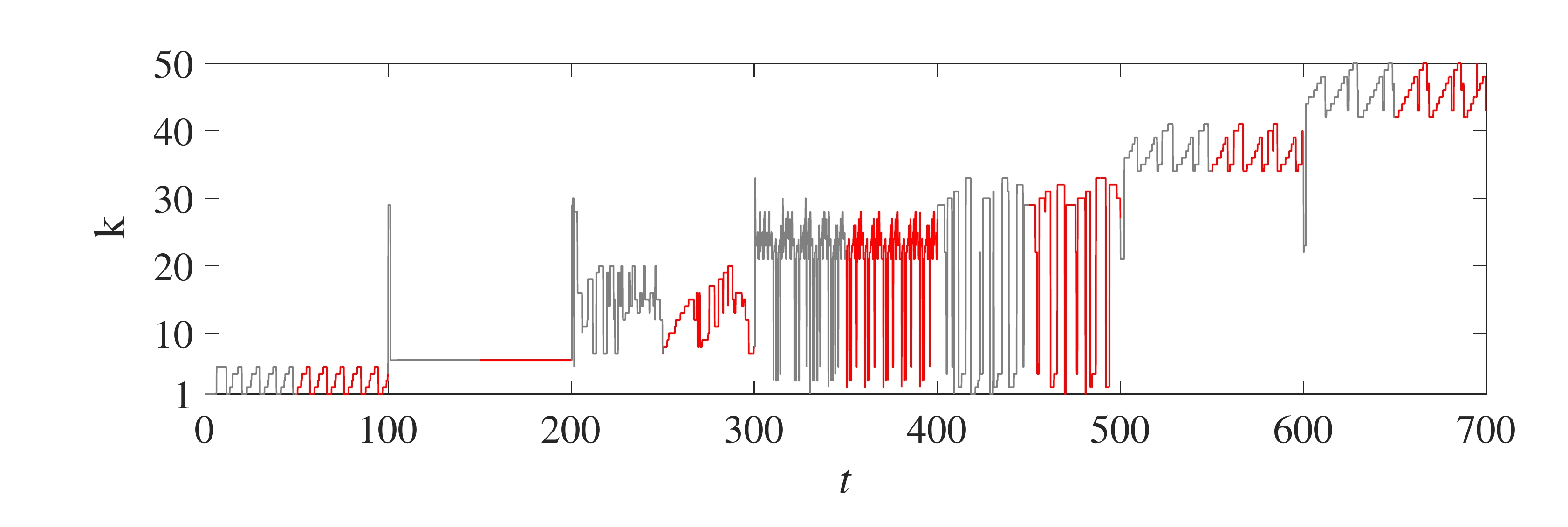}
	\caption{Cluster affiliation $k$ 
        of the fluidic pinball simulation as a  function of time $t$. 
        The applicable control laws are described in table~\ref{Tab:Pinball:Simulation}
        and the corresponding centroid $\bm{c}_k$ is depicted in figure~\ref{Fig:Pinball:Centroids}.
        By construction, new cluster indices are a monotonously increasing function of time, 
        neglecting the initial actuation transients.
        Gray curve sections correspond to transients 
        while red sections mark the posttransient data taken for the analysis.
}
\label{Fig:Pinball:SnapshotCluster}
\end{figure}
The cluster affiliation as a function of time 
is shown in figure \ref{Fig:Pinball:SnapshotCluster}.
This affiliation illustrates the role of each centroid (figure \ref{Fig:Pinball:Centroids})
in the posttransient behaviors of control laws $I$--$VII$ of 
Table \ref{Tab:Pinball:Simulation}.
Clusters 1--5 describe unforced vortex shedding of phase $I$ 
(see figure \ref{Fig:Pinball:Centroids}.
We emphasize that the centroids are only meant to resolve converged attractor data
and that the presentation of the transient dynamics shall only indicate the transient times.
The stabilized boat tailing of phase $II$ is described by a single centroid $k=6$.
Base bleed (phase $III$) is resolved by the new centroids 7--20,
indicating that base bleed leads to significantly different vortex shedding structures.
High-frequency (phase $IV$) and low-frequency forcing (phase $V$) 
can be resolved with the centroids of unforced vortex shedding and of base-bleed actuation.
Low-frequency forcing leads to a wider wake, like base bleed, and has more overlap with base-bleed dynamics.
The new centroids 34--41 resolve the positive Magnus effect (phase $VI$)
while centroids 42--50 the negative Magnus effect (phase $VII$).
As physical intuition suggests, the strong deflected wake has no overlap 
with forced states which are symmetric or statistically symmetric.

\begin{figure}
	\centering
	\includegraphics[width=80mm]{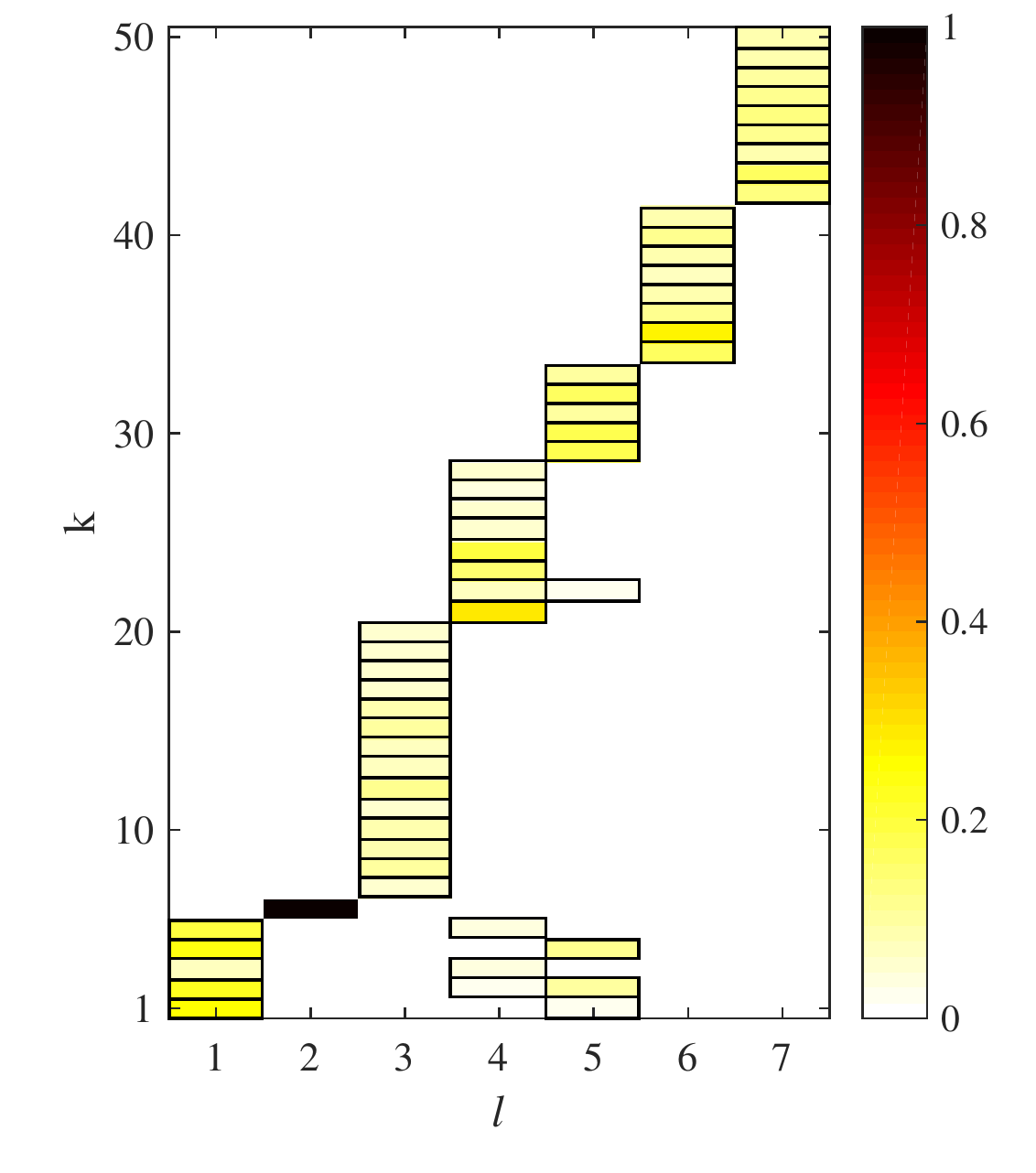}
	\caption{Population probabilities of the clusters 
        for the seven fluidic pinball phases (table \ref{Tab:Pinball:Simulation}).
        The abscissa denotes the phase $l$ and the ordinate the cluster index $k$.
        Each box represents a non-vanishing probability.
        The yellow to red tones indicate the population probability  $P_k^l$ from \eqref{Eqn:Frequency}.
        Probabilities below 1\% are kept white.
        Note that steady boat tailing ($l=2$) is represented by a single centroid.
}
\label{Fig:Pinball:Frequencies}            
\end{figure}
Figure~\ref{Fig:Pinball:Frequencies} displays the population 
of each cluster by the seven different dynamics.
Probabilities below 1\% are kept white.
The probability distributions for each attractor are far from being uniform.
This behavior is different from single attractor cluster analyses, 
where all clusters are generally populated~\citep{Kaiser2014jfm}.
In contrast, considering the whole simulation results 
in the total probability $P_k = P_k^1 + \ldots + P_k^7$
which is nonzero for any cluster $k$. 
Note that one cluster may be populated by different attractor dynamics.

\subsection{Metric of attractor overlap}
\label{Sec:Pinball:Metric}
 
\begin{figure}
	\centering
  \includegraphics[height=4cm]{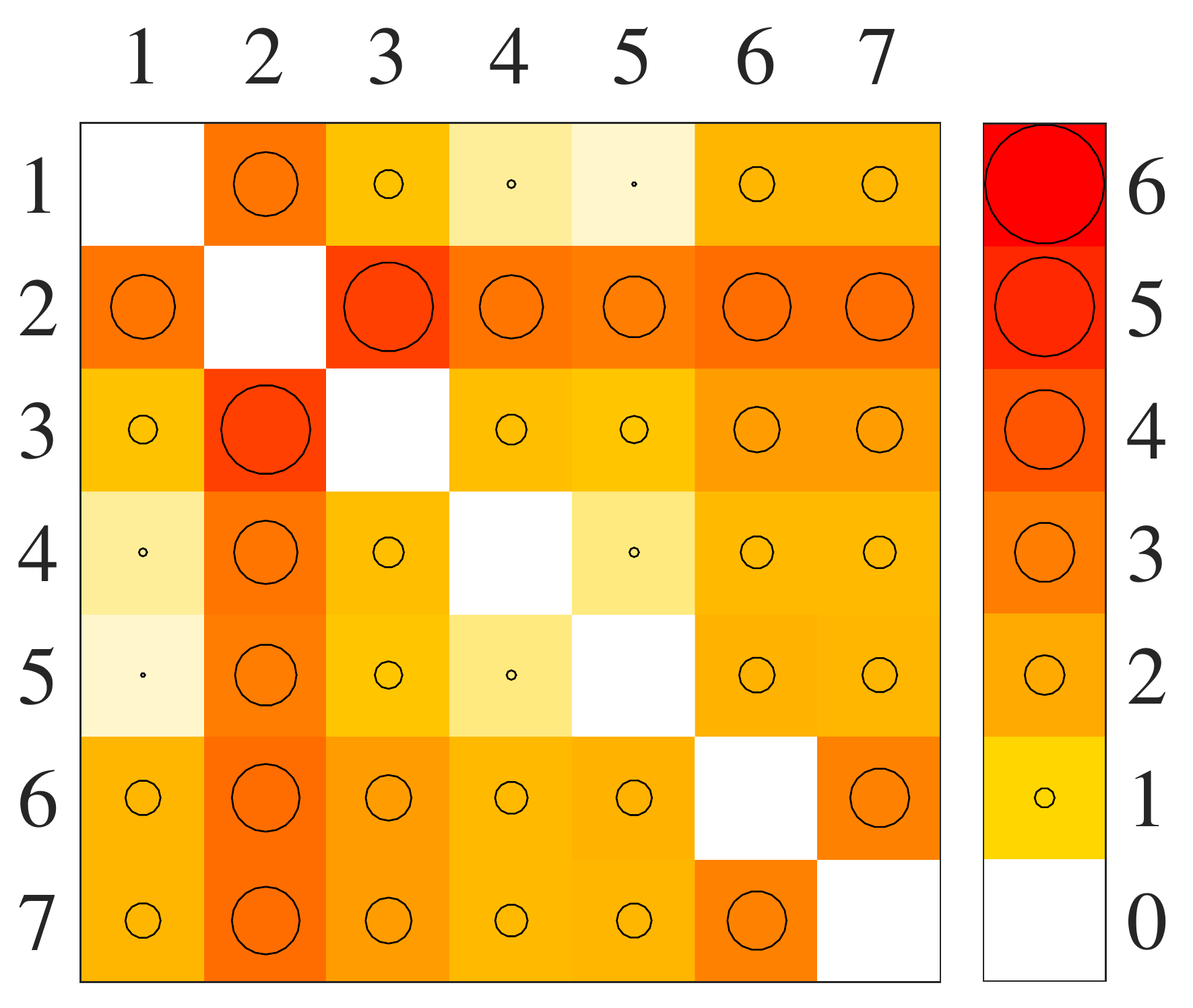}
	\caption{Metric of attractor overlap \eqref{Eqn:MAO:Centroids}
           for the seven control laws of the fluidic pinball (table \ref{Tab:Pinball:Simulation}). 
           The value of $D_{\rm geom} \left ( \vec{P}^l, \vec{P}^n  \right)$
is shown in the $l$th row and $n$th column
as color code from white to red 
and by the size of the circle as illustrated by the caption on the right.
}
\label{Fig:Pinball:GeometricMetric}
\end{figure}

The metric of attractor overlap for the seven operating conditions
is displayed in figure \ref{Fig:Pinball:GeometricMetric}.
The positive (semi)definiteness implies the vanishing elements in the diagonal
and non-negative values elsewhere.
The symmetry condition leads to the corresponding  symmetry of the matrix.
The values of the metric
are based  on the Hilbert space norm  \eqref{Eqn:MAO:Snapshots}
and an averaging process over the two attractors \eqref{Eqn:DistanceSnapshots}.
Hence, the fluctuation level, i.e., twice the fluctuation energy $2 \> \rm TKE$, 
represents a natural reference value for the square of the metric.
Values which are at least one order of magnitude smaller correspond to similar attractors.
For the unforced fluidic pinball configuration, 
the fluctuation level is $2 \> \rm TKE = 17.80$.
Hence, a MAO value of $\sqrt{2 \> \rm TKE}=4.22$ can be considered as reference scale for closeness.

Natural vortex shedding ($l=1$) is seen to be close 
to high- and low-frequency forcing ($l=4,5$),
as the metric is a tiny value of the maximum (small circles).
This closeness is corroborated by the probability distributions 
\ref{Fig:Pinball:Frequencies}: the three states share joint clusters.
In contrast, the wake stabilized with boat tailing
shares no centroids with the other six operating conditions
and has a large MAO distance to them.
The values are comparable or exceed the reference scale.
Similarly, 
the positive and negative Magnus effect ($l=6,7$) 
have a large distance to each other, 
as expected from the opposite deflections of the wake in
 figure \ref{Fig:Pinball:Simulation} 
and the empty overlap of both attractors,
i.e., no shared clusters in figure \ref{Fig:Pinball:Frequencies}.

\subsection{Proximity map}
\label{Sec:Pinball:Proximity}

\begin{figure}
	\centering
	\includegraphics[width=0.8\textwidth]{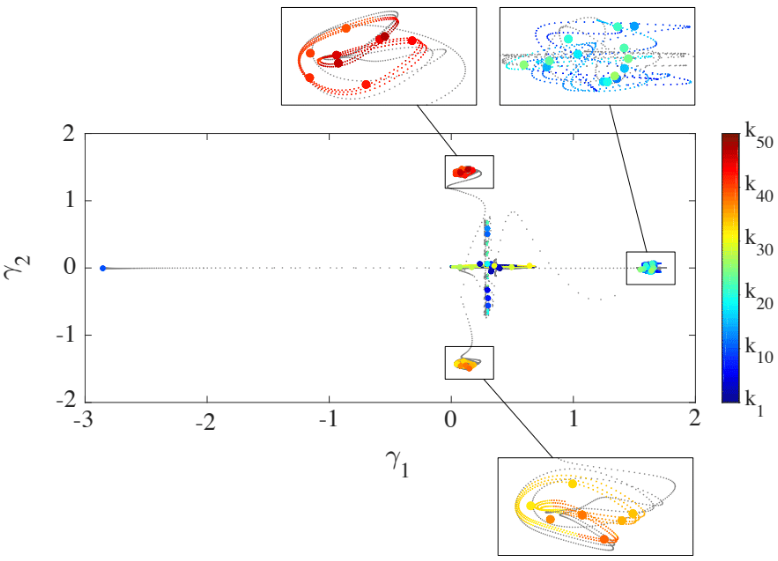}
	\caption{Proximity map of the fluidic pinball 
                 showing the centroids (large circles) together with snapshots (small circles). 
                 Snapshots from transients are  displayed as gray dots, 
while the data for the analysis is color-coded by cluster affiliation.}
\label{Fig:Pinball:CentroidProximity}
\end{figure}
Figure~\ref{Fig:Pinball:CentroidProximity}
 displays all centroids and all snapshots in a proximity map
following \S~\ref{Sec:Method:Proximity}.
Phase $I$---$V$ attractors are located on the $\gamma_2 = 0$ line.
These flows are statistically symmetric with respect to the $x$-axis 
and have, correspondingly, vanishing average lift.
Phase $VI$ and $VII$ correspond to positive and negative Magnus effects
associated with negative and positive average lift, respectively.
These phases are mirror-symmetrically located in the lower and upper region.
The second feature coordinate $\gamma_2$ is clearly related to averaged lift.
The lift values of all phases are displayed in figure \ref{Fig:Pinball:Simulation}.

The boat-tailed (phase $II$) and base-bleed dynamics (phase $III$) 
represent the minimal and maximum drag states of considered dynamics,
referring again to figure \ref{Fig:Pinball:Simulation}.
These attractors are on the leftmost and rightmost sides near the $\gamma_1$--axis, respectively.
The other attractors have similar  drag and have  $\gamma_1\approx 0$.
Summarizing, the first feature coordinate is strongly correlated to drag.
It should be noted that the automatically determined feature coordinates
are strongly linked to aerodynamic forces although the forces did not enter the multi-dimensional scaling analysis.

\begin{figure}
	\centering
   \begin{tabular}{ll}
		(a) & (b)  \\
		\includegraphics[width=0.45\textwidth]{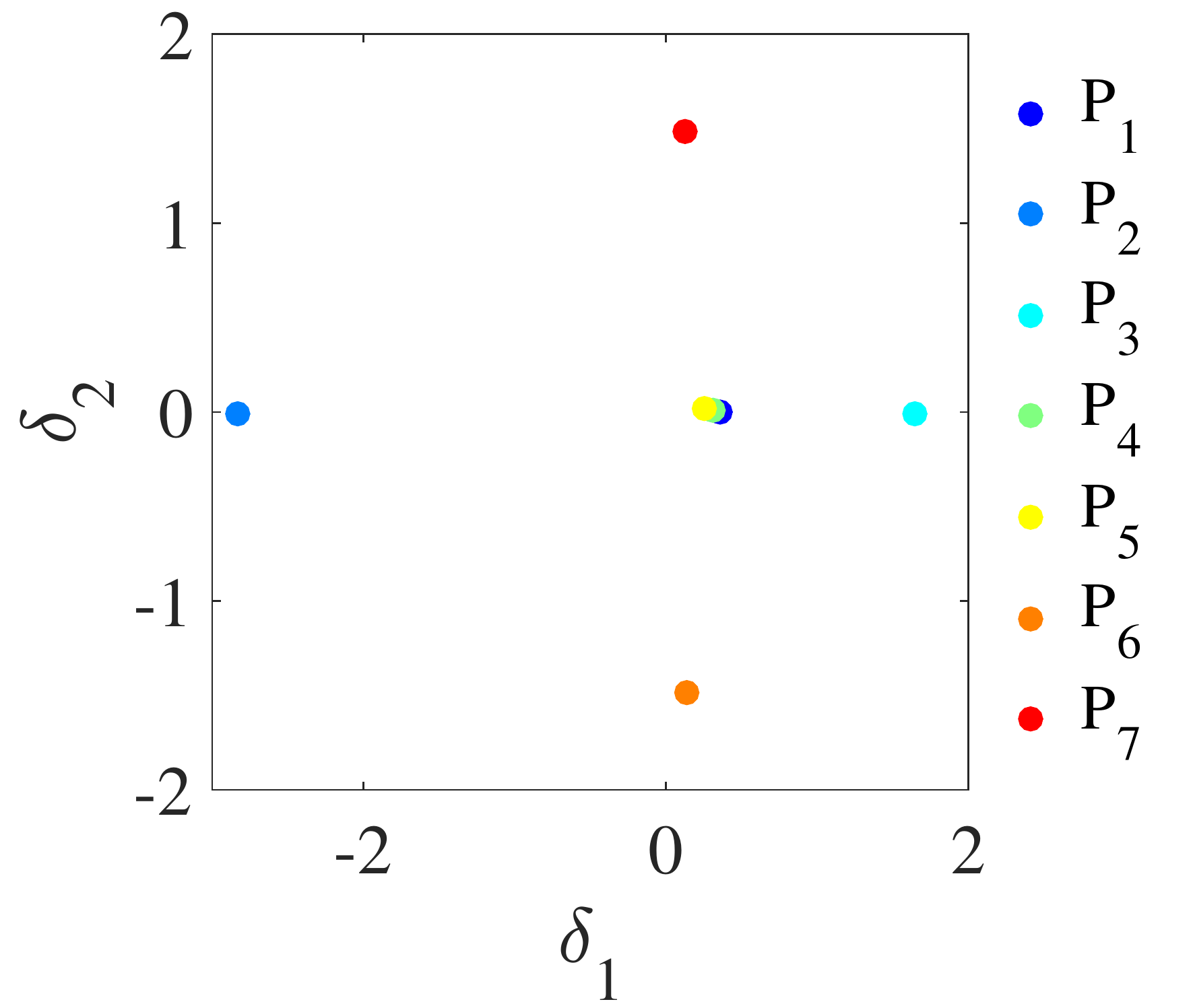} &
		\includegraphics[width=0.45\textwidth]{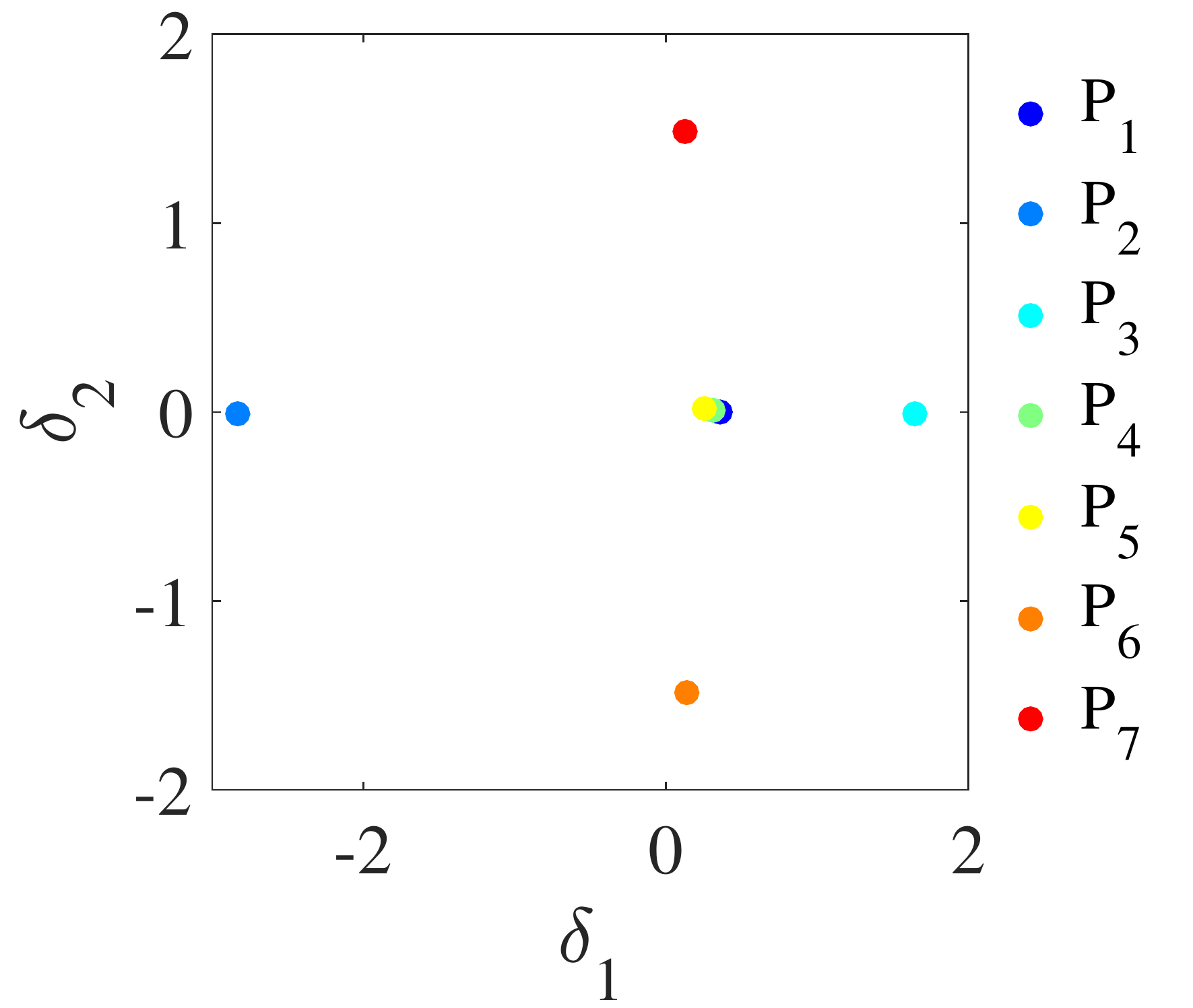}
	\end{tabular}
	\caption{Attractor proximity map of the fluidic pinball:
                 (a) based on the snapshot-based MAO \eqref{Eqn:MAO:Snapshots} and
                 (b) on a  cluster-based  MAO \eqref{Eqn:MAO:Centroids} with $K=50$ clusters.
The points $P_1 \ldots P_7$ correspond to the seven control laws in table \ref{Tab:Pinball:Simulation}.}
\label{Fig:Pinball:AttractorProximity}
\end{figure}
The proximity map for the attractors based on  MAO
is  depicted in figure \ref{Fig:Pinball:AttractorProximity},
both, for the snapshot-based definition (a) and the cluster-based estimate (b).
By construction, 
the distances between the feature vectors of the seven attractors
are similar the MAO metric values displayed in figure \ref{Fig:Pinball:GeometricMetric}.
are similar to 
Both maps are virtually indistinguishable, thus showing the accuracy of the cluster-based approximation.
Varying the cluster resolution in a wide range $K \in \{25,50,100\}$ 
corroborates the robustness of the cluster-based MAO: For all values,
the proximity maps  are an excellent approximation of the snapshot-based version.

The feature coordinate $\delta_1$ correlates with the drag and
$\delta_2$ with the average lift---like in  figure \ref{Fig:Pinball:CentroidProximity}.
Even the numerical values of the feature coordinates 
associated with the same operating conditions are similar.
and the attractor proximity map is an aggregate of the snapshots and centroids.
This quantitative similarity indicates 
the robustness of the multi-dimensional scaling 
for the construction of proximity maps.

\section{Comparison of turbulent boundary layers at different wall actuation}
\label{Sec:TBL:Results}

We analyze the three-dimensional flow data 
generated by 38 simulations of the actuated turbulent boundary layer 
using the cluster-based MAO methodology. 
In \S~\ref{Sec:TBL:Clustering} 
we present the cluster analysis of the 38 attractors. 
In \S~\ref{Sec:TBL:Metric} 
and \S~\ref{Sec:TBL:Proximity} 
the MAO analysis and proximity maps for snapshots, centroids and attractors
are illustrated.

\subsection{Cluster analysis}
\label{Sec:TBL:Clustering}
\begin{figure}
	\centering
        \begin{tikzpicture}
          \tikzstyle{every node}=[font=\small]
          \tikzset{>=latex}
          \def\picwidth{10cm}
          \node[anchor=south west,inner sep=0] (image) at (0,0) {
            \includegraphics[width=\picwidth]{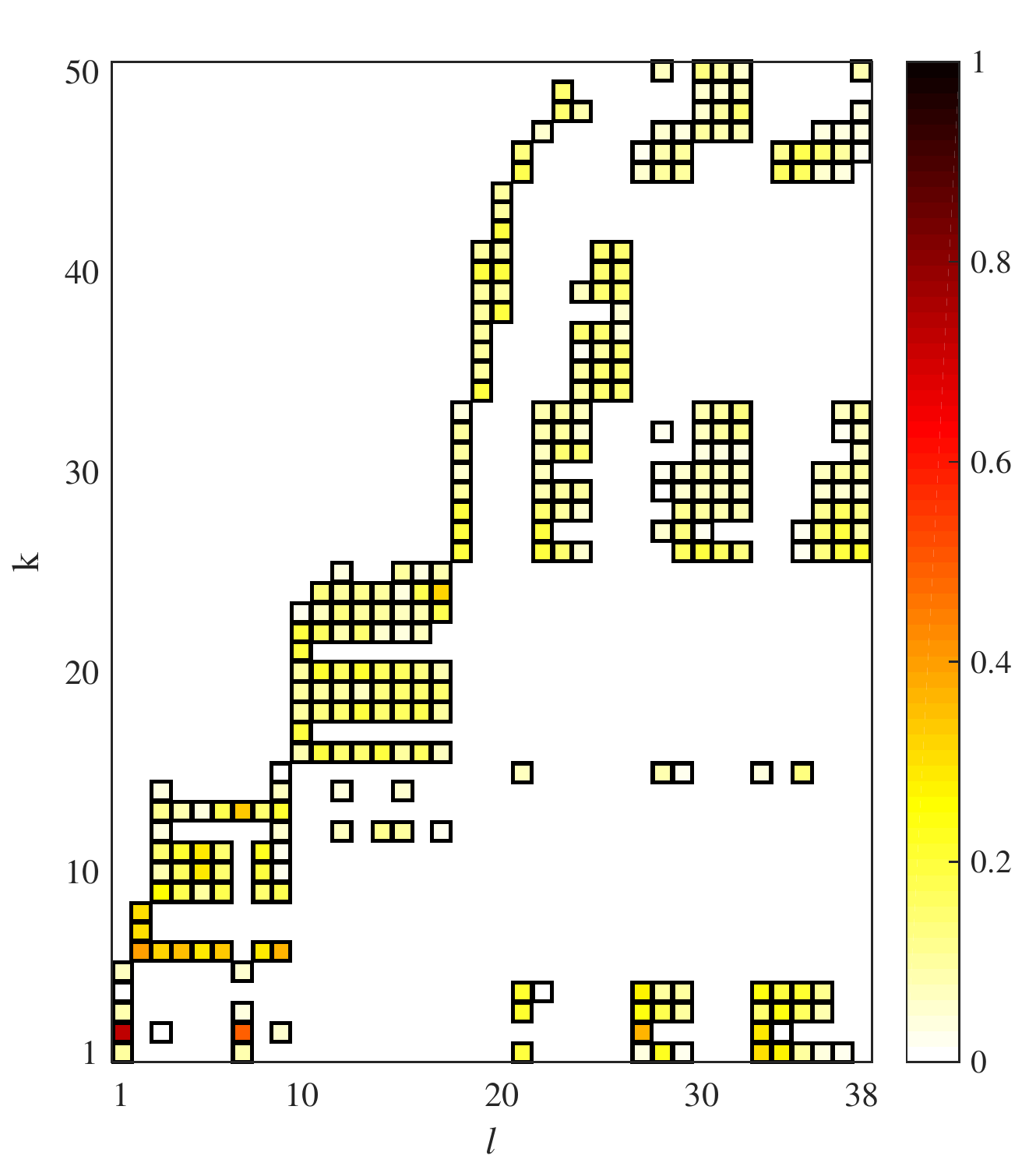}
          };
          \draw (0.128*\picwidth,0.1*\picwidth) -- (0.128*\picwidth,1.0652*\picwidth);
          \draw (0.2821*\picwidth,0.1*\picwidth) -- (0.2821*\picwidth,1.0652*\picwidth);
          \draw (0.4363*\picwidth,0.1*\picwidth) -- (0.4363*\picwidth,1.0652*\picwidth);
          \begin{scope}[x={(image.south east)},y={(image.north west)},shift=(image.south west)]
            \begin{axis}[ylabel=$A^+$,axis lines=left, xtick=\empty, ytick=\empty, 
                         xshift=0.11*\picwidth,yshift=1.1*\picwidth,width=0.73*\picwidth,height=0.05*\picwidth,
                         xmin=-0.5,xmax=37.5,ymin=0.0,ymax=70, x axis line style={-},
                         scale only axis=true]
              \addplot[ybar,line width=1.0pt,fill=red,bar width=3.0] coordinates {
                (1,30)
                (2,21)
                (3,30)
                (4,45)
                (5,30)
                (6,14)
                (7,38)
                (8,28)

                (9,30)
                (10,22)
                (11,21)
                (12,30)
                (13,30)
                (14,36)
                (15,64)
                (16,48)

                (17,10)
                (18,30)
                (19,50)
                (20,10)
                (21,20)
                (22,30)
                (23,40)
                (24,50)
                (25,60)
                (26,10)
                (27,20)
                (28,30)
                (29,40)
                (30,50)
                (31,60)
                (32,10)
                (33,20)
                (34,30)
                (35,40)
                (36,50)
                (37,60)
              };
            \end{axis}
            \begin{axis}[ylabel=$T^+$,axis lines=left, xtick=\empty, ytick=\empty, 
                         xshift=0.11*\picwidth,yshift=1.2*\picwidth,width=0.73*\picwidth,height=0.05*\picwidth,
                         xmin=-0.5,xmax=37.5,ymin=0.0,ymax=130, x axis line style={-},
                         scale only axis=true]
              \addplot[ybar,line width=1pt,fill=yellow,bar width=3.0] coordinates {
                (1,20)
                (2,30)
                (3,40)
                (4,50)
                (5,60)
                (6,70)
                (7,70)
                (8,100)

                (9,20)
                (10,30)
                (11,40)
                (12,40)
                (13,60)
                (14,70)
                (15,70)
                (16,100)

                (17,20)
                (18,20)
                (19,20)
                (20,40)
                (21,40)
                (22,40)
                (23,40)
                (24,40)
                (25,40)
                (26,80)
                (27,80)
                (28,80)
                (29,80)
                (30,80)
                (31,80)
                (32,120)
                (33,120)
                (34,120)
                (35,120)
                (36,120)
                (37,120)
              };
            \end{axis}
          \end{scope}
        \end{tikzpicture}
	\caption{Population probabilities of the clusters 
        for the 38  turbulent boundary layer simulations (see table \ref{Tab:TBL:Simulation}).
        The abscissa denotes the data set $l$ and the ordinate the cluster index $k$.
        Each box represents a non-vanishing probability.
        The yellow to red tones indicate the population probability or relative frequency $P_k^l$ from \eqref{Eqn:Frequency}.
        Probabilities below 1\% are kept white.
        The actuation wavelength is indicated by the three vertical line  separating simulations
        with no actuation, $\lambda^+=200$, $\lambda^+=500$ and $\lambda^+=1000$ from left to right.
        The actuation amplitude $A^+$ and time $T^+$ are shown in the top caption.}
\label{Fig:TBL:Frequencies}            
\end{figure}
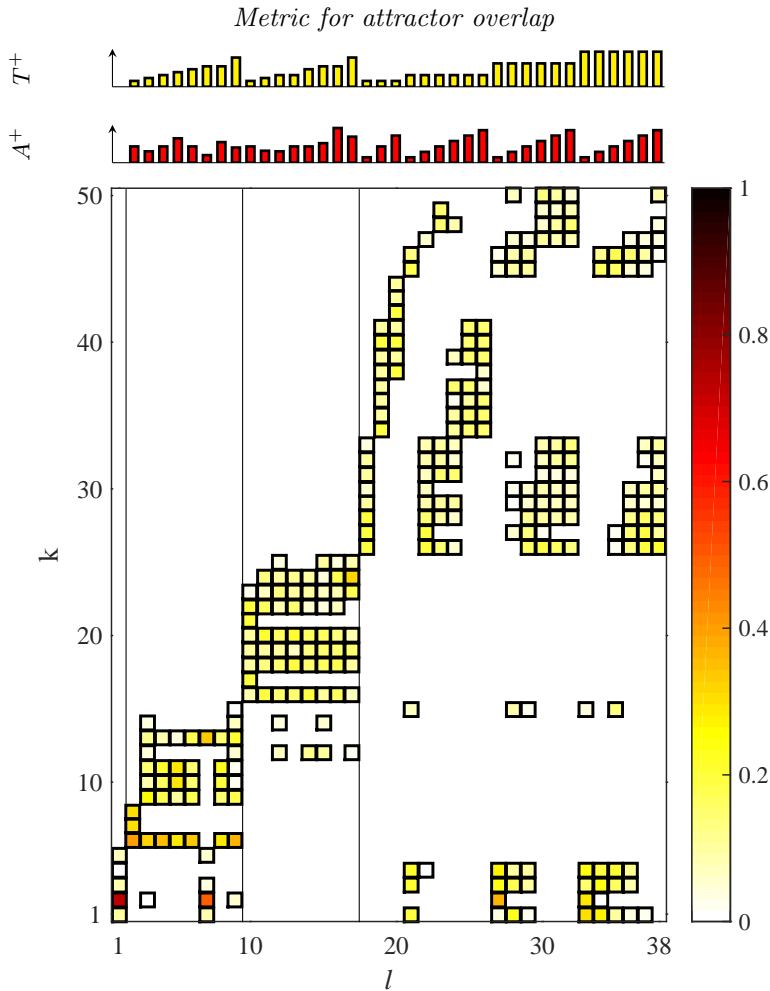
Figure \ref{Fig:TBL:Frequencies} displays the probability distribution of each attractor 
to pass through a given cluster. 
Again, probabilities below 1\% are kept white. 
All attractors have pair-wise different cluster compositions. 
The unforced reference $l=1$ occupies the first five clusters.
The attractors $l=2,\ldots,9$ corresponding to a spanwise wavelength $\lambda^⁺=200$
populate clusters $k=5,\ldots,16$.
The attractor $l=7$ with low actuation amplitude shares four of five clusters 
with the unforced flow and populates one new cluster $k=14$.
The simulations with a spanwise wavelength $\lambda^+=500$ ($l=10,\ldots,17$)
pass through the new clusters $k=18,\ldots,26$.
These attractors have no overlap with the unforced reference and share few states 
with the previous group at low-wavelength actuation. 
The long-wavelength actuation ($\lambda^+ = 1000$) 
populates the new clusters $l=27,\ldots,50$.
This group shares only one cluster $(l=16)$ with the medium-wavelength group
but may have significant overlap with the unforced reference.
This is particularly true for low-amplitude ($A^+=10$ ) actuations at $l=21$, $27$, and $33$.
It should be noted that the MAO methodology displays 
the overlap between attractors in a computer- and human-interpretable form.

\subsection{Metric of attractor overlap}
\begin{figure}
	\centering
  \includegraphics[height=8cm]{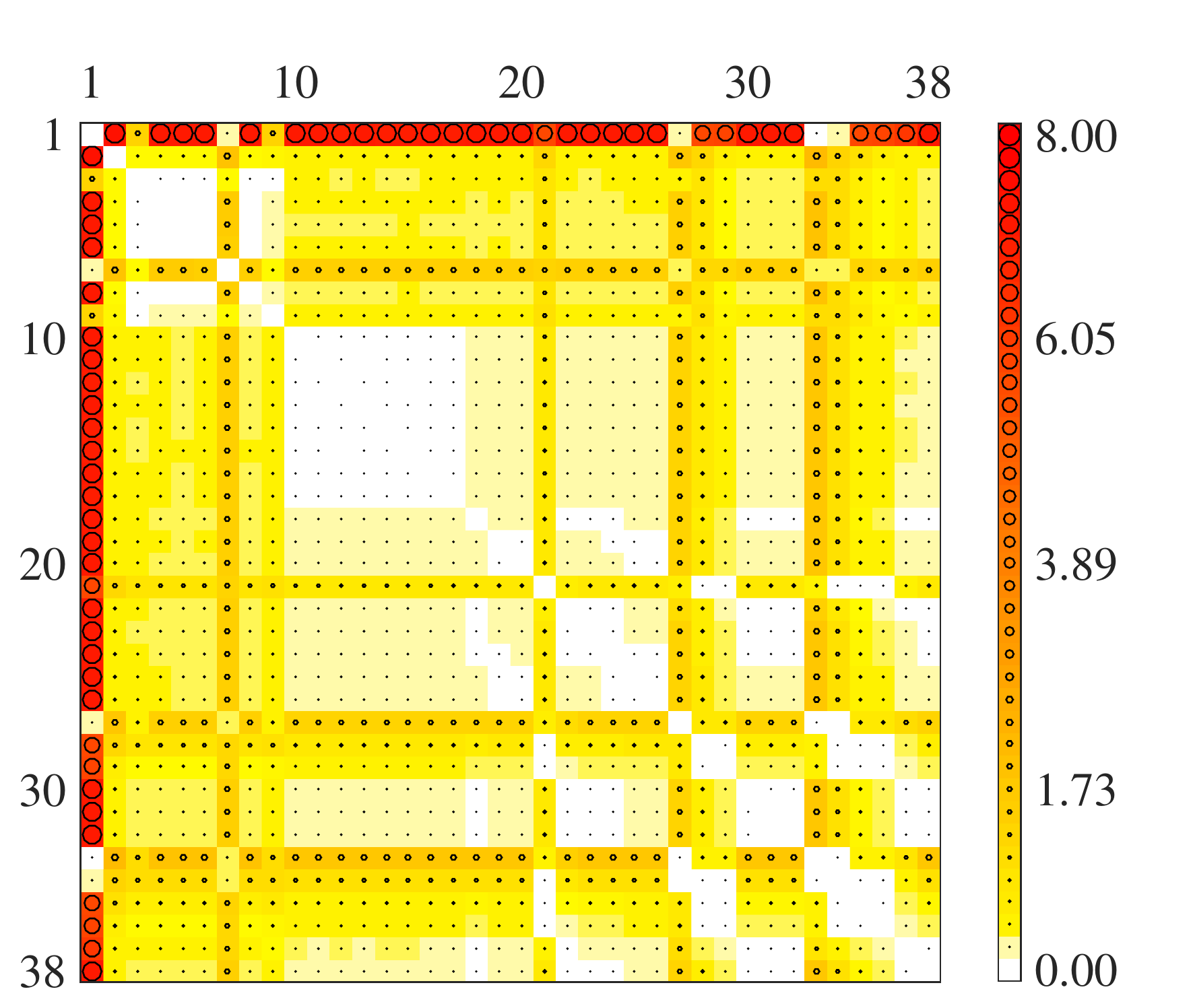}
 	\caption{Metric of attractor overlap \eqref{Eqn:MAO:Centroids} of the turbulent boundary layer
           with 38 control laws (see table \ref{Tab:TBL:Simulation}). 
           The visualization is analogous to figure \ref{Fig:Pinball:GeometricMetric}.}
\label{Fig:GeometricMetric_TBL}
\end{figure}
\label{Sec:TBL:Metric}
Figure \ref{Fig:GeometricMetric_TBL} displays the geometric distance 
between the attractors based on MAO. 
Evidently the diagonal vanishes by definition.
In complete analogy to the fluidic pinball,
the reference for a large scale can be taken 
to be the fluctuation amplitude $\sqrt{2 \> \rm  TKE} = 0.8767$
of the unforced flow $(A=0)$, 
noting that actuation may increase this amplitude by one order of magnitude.
Most of metric values are in this range or lower.

The unforced reference $l=1$  is seen to be very different from most other attractors. 
Attractors 2 to 6 are very close to each other, 
which can be expected because of their same actuation wavelength $\lambda^+=200$. 
Again, attractors 10 to 17 corresponding to wavelength $\lambda^+=500$
display significant similarity, i.e.\ a low MAO. 
The group of actuated cases with large wavelength $\lambda^+=1000$ show less similarity
which is consistent with the discussed overlap from the previous figure.
Attractors 7, 21, 27, and 33 have  low actuation amplitude, 
share similar characteristics with the unforced reference
as may be expected. 

\subsection{Proximity map}
\label{Sec:TBL:Proximity}
\begin{figure}
	\centering
	\includegraphics[width=0.5\textwidth]{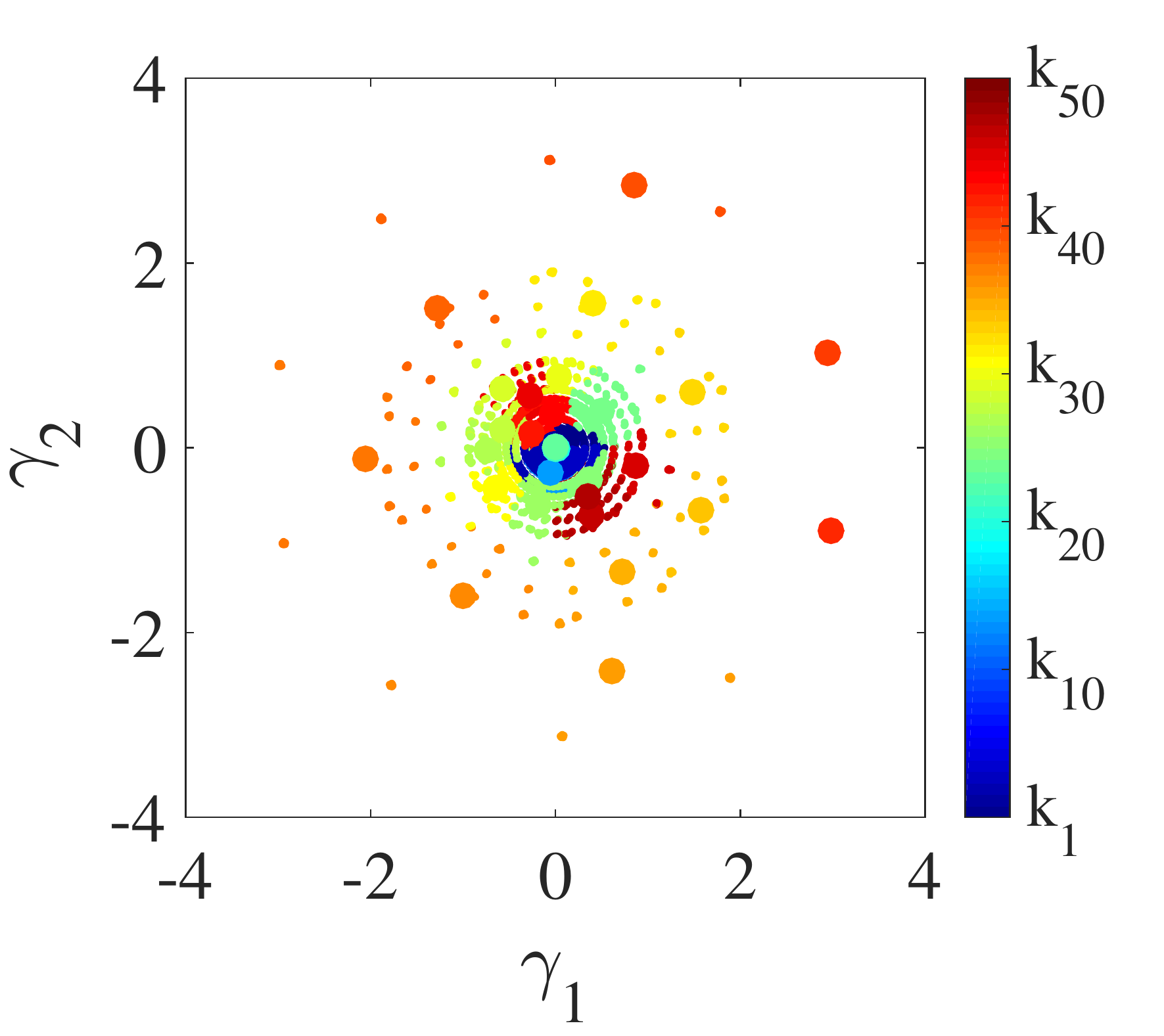}
	\caption{Proximity map of the actuated turbulent boundary layer.
               The figure displays the the centroids (large circles) together with snapshots (small circles). 
               The data for the analysis is color-coded by cluster affiliation.}
\label{Fig:CentroidProximity}
\end{figure}
Figure \ref{Fig:CentroidProximity} 
visualizes the  distance between the centroids and snapshots. 
Actuations at low wavelengths (small $k$) are close to the origin
while large wavelength actuation (large $k$) 
may populate circular regions in the periphery.
A closer analysis reveals that $\gamma_1$, $\gamma_2$ correspond to
the first POD mode amplitudes $a_1$, $a_2$ computed from all snapshot data together.
These POD mode amplitudes are nearly periodic, particularly for large actuation amplitudes.

\begin{figure}
	\centering
\begin{tabular}{ll}
	(a) & (b) \\
\includegraphics[width=0.45\textwidth]{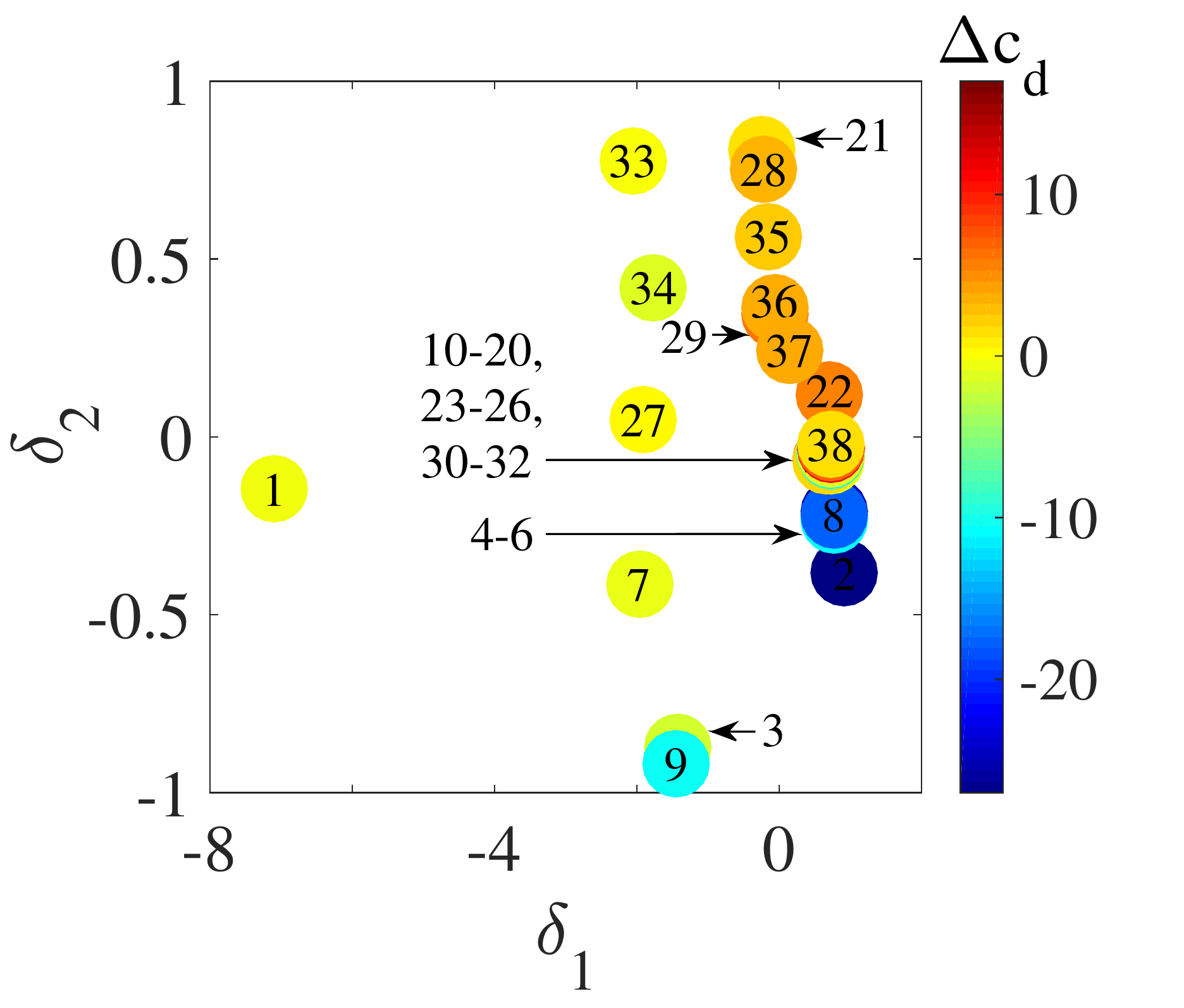} &
\includegraphics[width=0.45\textwidth]{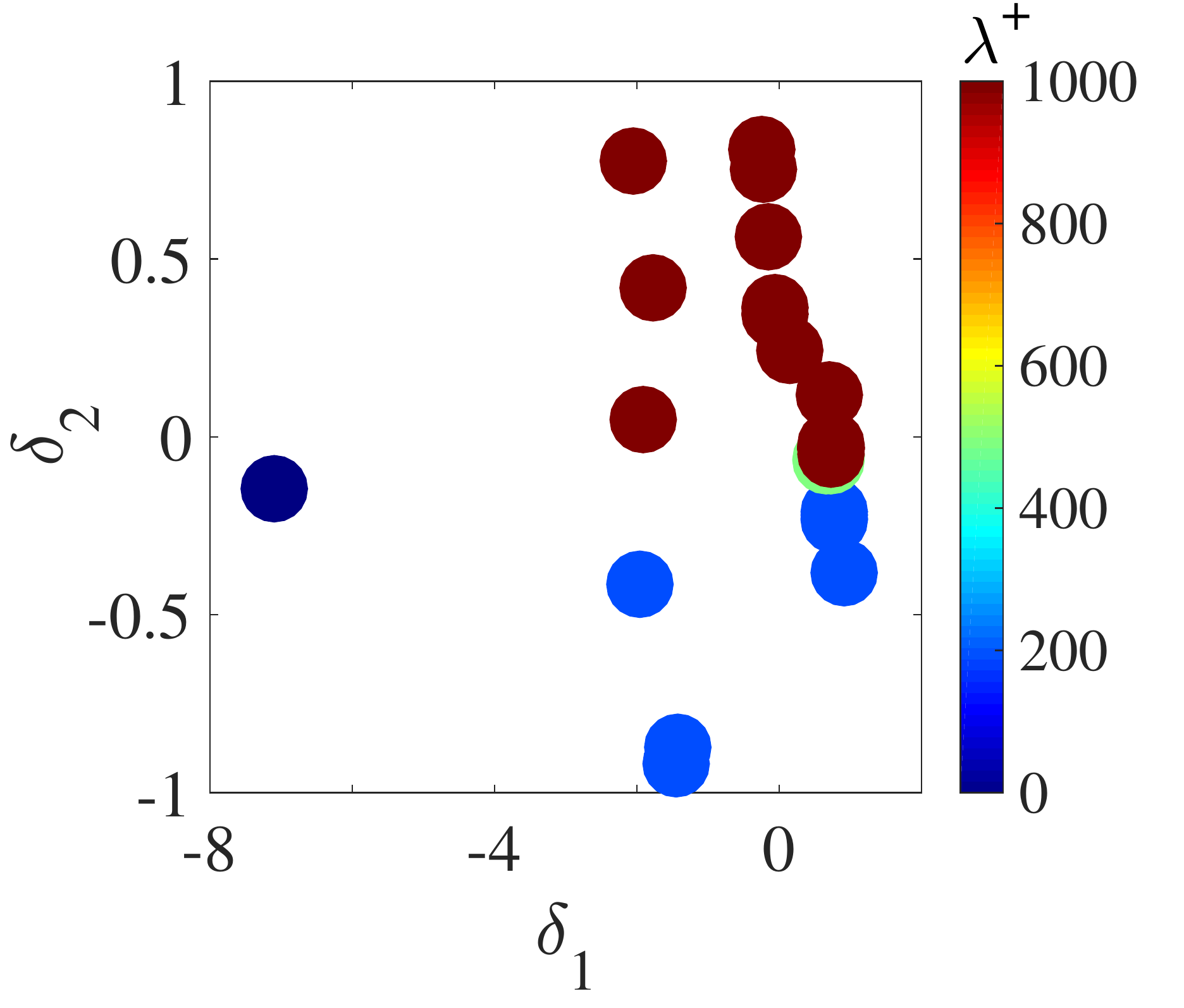}
\end{tabular}

	\caption{Attractor proximity map of the turbulent boundary layer simulations (see table~\ref{Tab:TBL:Simulation}): 
		(a) color-coded with relative drag reduction  $\Delta c_D$;
		(b) color-coded with wavelength $\lambda^+$.
The actuation parameters of each symbol is indicated in subfigure (a)
by a number corresponding to 
the index from table \ref{Tab:TBL:Simulation}.   
}
\label{Fig:ProximityMaps_DR_L}
\end{figure}
Figure \ref{Fig:ProximityMaps_DR_L} displays the proximity map
of the attractors based on  MAO.  
From Figure \ref{Fig:ProximityMaps_DR_L} (a)
the first feature coordinate $\delta_1$ 
is linked to the drag reduction for the attractors:
For $\delta_1 < 0$  actuation has little effect on drag 
but for $\delta_1 \ge 0$ these changes are significant---positive  for $\delta_2>0$
and negative for $\delta_2<0$.
From Figure \ref{Fig:ProximityMaps_DR_L} (b), $\delta_2$ correlates 
with the wavelength of the traveling wave. 
The large  wavelengths are located in the top half of the plot.
As seen in the fluidic pinball, 
the MAO-based attractor proximity map
extracts the aerodynamics and actuation properties 
without directly including information about them.
\section{Conclusions and outlook}
\label{Sec:Conclusions}

We have proposed a novel quantitative measure, 
termed \emph{Metric of Attractor Overlap} (MAO),
for the similarity of attractor data.
This measure generalizes the Hilbert-space metric for two velocity fields
to a metric between two sets with snapshot data.
A refined comparison analysis includes
proximity maps
and coarse-graining with clustering.

Our proposed comparison methodology has five discriminating features.
Firstly, the difference between two attractors is quantified in a single, non-negative number 
encapsulating all resolved coherent structures.
Secondly, the cluster probability distribution of each attractor
identifies shared and disjoint flow states via a manageable number of centroids.
Thirdly, the metric of attractor overlap 
enables the automatic creation of a proximity map from a myriad of attractors. 
Neighboring feature points represent attractors with similar flow states,
largely separated points indicate attractors with different physics.
Fourthly, the approach is compatible with statistical post-processing:
The centroids and their probabilities define the mean flow 
and---in principle---any higher-order moment,
like Reynolds stress or fluctuation energy.
Even POD modes can be derived from the centroids.
Finally, the comparison methodology can be automatized 
and has little subjective bias.
One does not need to decide in advance on a cost function, such as drag power, 
 a frequency filter, such as low-pass filter,
  important flow features, such as vortices, just to name a few.

The metric of attractor overlap is only based on two decisions.
First, one needs to choose the flow state space, 
e.g.\ the velocity field in a domain $\Omega$.
Such a decision is unavoidable for any metric.
Second, a metric for two flow states is needed.
A $L^2(\Omega)$ Hilbert space norm appears as a natural candidate.
For the optional coarse-graining, 
 the number of clusters or, synonymously, the typical size of a cluster needs to be chosen.
The coarse-grained metric is found to quickly converge to the snapshot-based one 
with increasing number of clusters.

The comparison methodology has been applied to the fluidic pinball,
the flow around a cluster of three rotating cylinders.
The considered attractors include the unforced reference,
aerodynamic boat tailing, base bleed, symmetric high-frequency forcing,
symmetric low frequency forcing and a Magnus effect deflecting the wake to both sides.
The metric of attractor overlap confirms physical inspection of the seven  configurations.
First, boat tailing with complete wake suppression differs strongly from all other states. 
Second, both Magnus effects also strongly differ from all other states.
Third, unforced vortex shedding and wakes 
manipulated by base bleed as well as
high-frequency  and low-frequency forcing have a significant overlap of shared clusters,
i.e.\ are similar.

Particularly noteworthy are the proximity maps of the snapshots 
and the cluster-based attractors.
The trajectory through all seven configurations displays the dynamics very clearly (see figure~\ref{Fig:Pinball:Simulation}).
Surprisingly,
the first and second feature coordinates resemble the drag and lift respectively (see figure~\ref{Fig:Pinball:AttractorProximity}).
The features of the snapshots
show the reduced drag by boat tailing (left)
and the increased drag by base bleed (right) and significant average lift by both Magnus forcing (top and bottom).
The proximity map of the attractor visually elucidates 
the discussed neighborhood relations in the first three feature coordinates.

The second application of MAO is an open-loop drag reduction study of a  turbulent boundary layer.
The operating conditions include the unforced reference 
and 37 actuations with spanwise traveling waves at different amplitudes and frequencies.
The spanwise wavelengths include
a small ($\lambda^+=200$), medium ($\lambda^+=500$), and large value ($\lambda^+=1000$).
The computational and observation domain are defined by the largest actuation wavelength.
The MAO analysis features several highlights.
First, the low amplitude actuations share centroids with the unforced reference
in agreement with physical intuition.
Second, actuations with the same spanwise wavelength show significant overlap.
Third, the feature map of the snapshots show low amplitude states near the center
and large amplitude actuations populate outer circular regions.
The feature coordinates are well aligned with POD mode amplitudes of the phase-averaged flow 
using actuation as a clockwork.
Finally, the first feature coordinate of the attractor proximity map
correlates well with the drag reduction---like for the fluidic pinball---
while the second coordinate depends on the  spanwise actuation wavelength.

The observed correlation between first feature coordinate 
and drag for two independent configurations  is surprising
as the input data does not contain information about drag.
In hindsight, this behavior may be explained by the strong correlation
between the time-averaged flow and the drag---both, for the wake and for the boundary layer. 

Evidently, the presented MAO comparison 
encourages  numerous other applications,
like a comparison between computed and experimental velocity fields
and a comparison between flow behavior under different actuations,
e.g. periodic forcing or machine learning control studies \citep{Duriez2016book}.
The analysis may also be applied to sensor data, e.g.\ a hot-wire rig,
instead of velocity fields.

Future research may significantly extend the MAO methodology.
The employed Hilbert space norm is a good initial choice
but highly sensitive to small mode deformations \citep{Noack2016jfm2}.
A small change in wavenumber gives rise to unphysically large differences in the norm.
Force-related feature vectors and manifold learning may be one remedy \citep{Loiseau2018jfm}.
Rigorous physics-based criteria for the cluster numbers need to be advanced.
The comparison may also be targeted towards single- or
multi-objective goals by generalizing the snapshot metric.
So far, the comparison only addressed ergodic properties of the attractor,
and the snapshot data only needed to be statistically representative and not time-resolved.
The temporal dynamics of different attractors 
may be compared using time-resolved snapshots
and corresponding Markov models \citep{Kaiser2014jfm}.

Summarizing,  a rational automated comparison method has been proposed 
which holds significant promise for future data assessments.
The authors are actively pushing this direction.

\section*{Acknowledgments}
\begin{acknowledgments}
This work is supported 
by a public grant overseen by the French National Research Agency (ANR) 
as part of the ``Investissement d’Avenir'' program, through the  ``iCODE Institute project'' funded by the IDEX Paris-Saclay, ANR-11-IDEX-0003-02,
by the ANR grants ``ACTIV\_ROAD'' and ``FlowCon'',
 by the Polish National
Center for Research and Development under the Grant No. PBS3/B9/34/2015,
and by the Bernd Noack Cybernetics Foundation.
Additional funds are provided by the German Research Foundation (DFG)
under grant numbers SE 2504/2-1 and SCHR 309/68-1.
Computing resources were provided by the High Performance Computing Center Stuttgart (HLRS) and by the J\"ulich Supercomputing Center (JSC)
within a large-scale project of the Gauss Center for Supercomputing (GSC).

EK gratefully acknowledges support by the ``Washington Research Foundation Fund for Innovation in Data-Intensive Discovery" and a Data Science Environments project award from the Gordon and Betty Moore Foundation (Award \#2013-10-29) and the Alfred P. Sloan Foundation (Award \#3835) to the University of Washington eScience Institute.

We appreciate valuable stimulating discussions with
Steven Brunton, Nathan Kutz, Themis Sapsis, and 
the French-German-Canadian-American pinball team: 
Guy Yoslan Cornejo-Maceda, Nan Deng,  Fran\c{c}ois Lussgeeyran, Robert Martinuzzi, Cedric Raibaudo,  Romain Rolland and Luc Pastur.
\end{acknowledgments}

\appendix
\section{Proper orthogonal decomposition versus clustering}
\label{Sec:POD}

The cluster analysis is compared with the proper orthogonal decomposition (POD)
from the same data in the same observation domain.
Each snapshot is expanded in terms of 
the mean flow $\vec{u}_0 (\vec{x})$ of all post-transient attractor data
and $N$ POD modes $\vec{u}_i(\vec{x})$, $i=1,\ldots,N$ 
and their corresponding amplitudes $a_i(t)$, $i=1,\ldots,N$:
\begin{equation}
\label{Eqn:POD}
\vec{u}(\vec{x},t) \approx \vec{u}_0 ( \vec{x} ) + \sum\limits_{i=1}^N a_i(t) \> \vec{u}_i(\vec{x}).
\end{equation}

\begin{figure}
	\centering
	\includegraphics[width=0.9\textwidth]{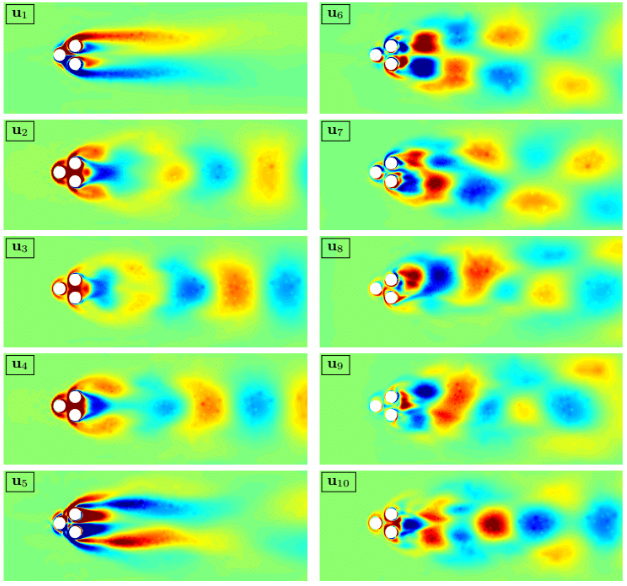}
	\caption{POD modes $\vec{u}_i(\vec{x})$, $i=1,\ldots,10$.}
\label{Fig:Pinball:PODModes}	
\end{figure}

\begin{figure}
	\centering
	\includegraphics[width=6cm, trim=0 3.5cm 0 0cm, clip=true]{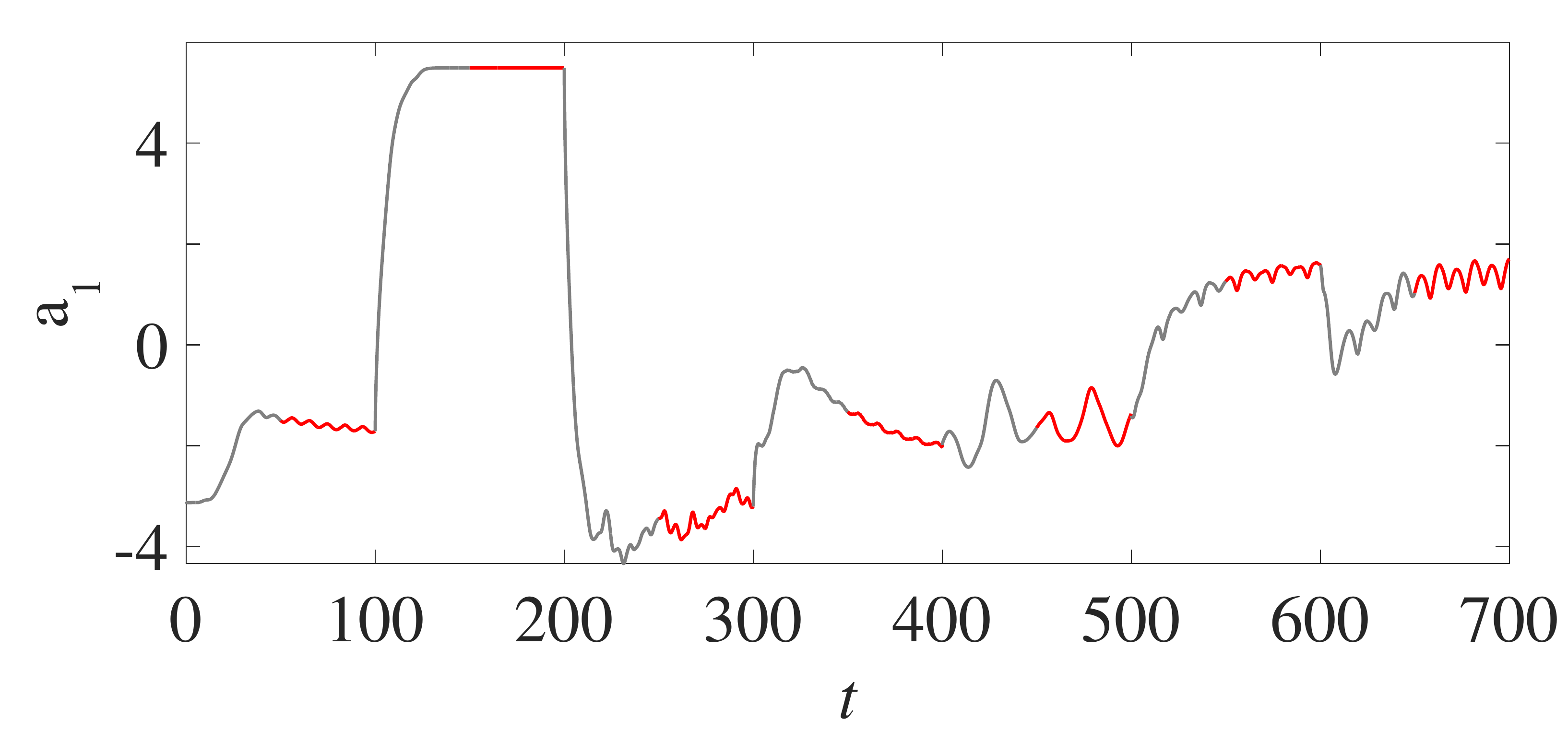}	
	\includegraphics[width=6cm, trim=0 3.5cm 0 0cm, clip=true]{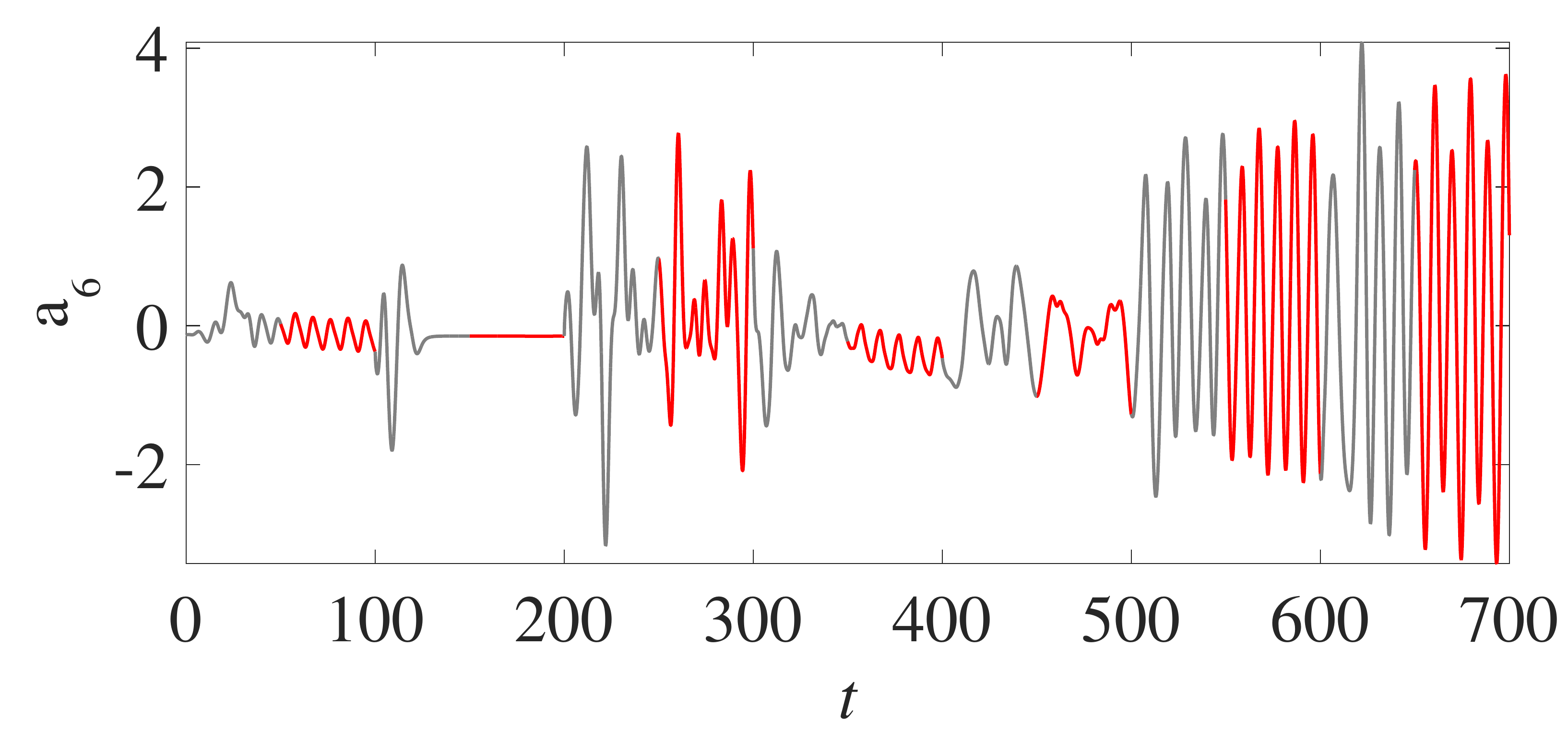}\\[1ex]
	\includegraphics[width=6cm, trim=0 3.5cm 0 0cm, clip=true]{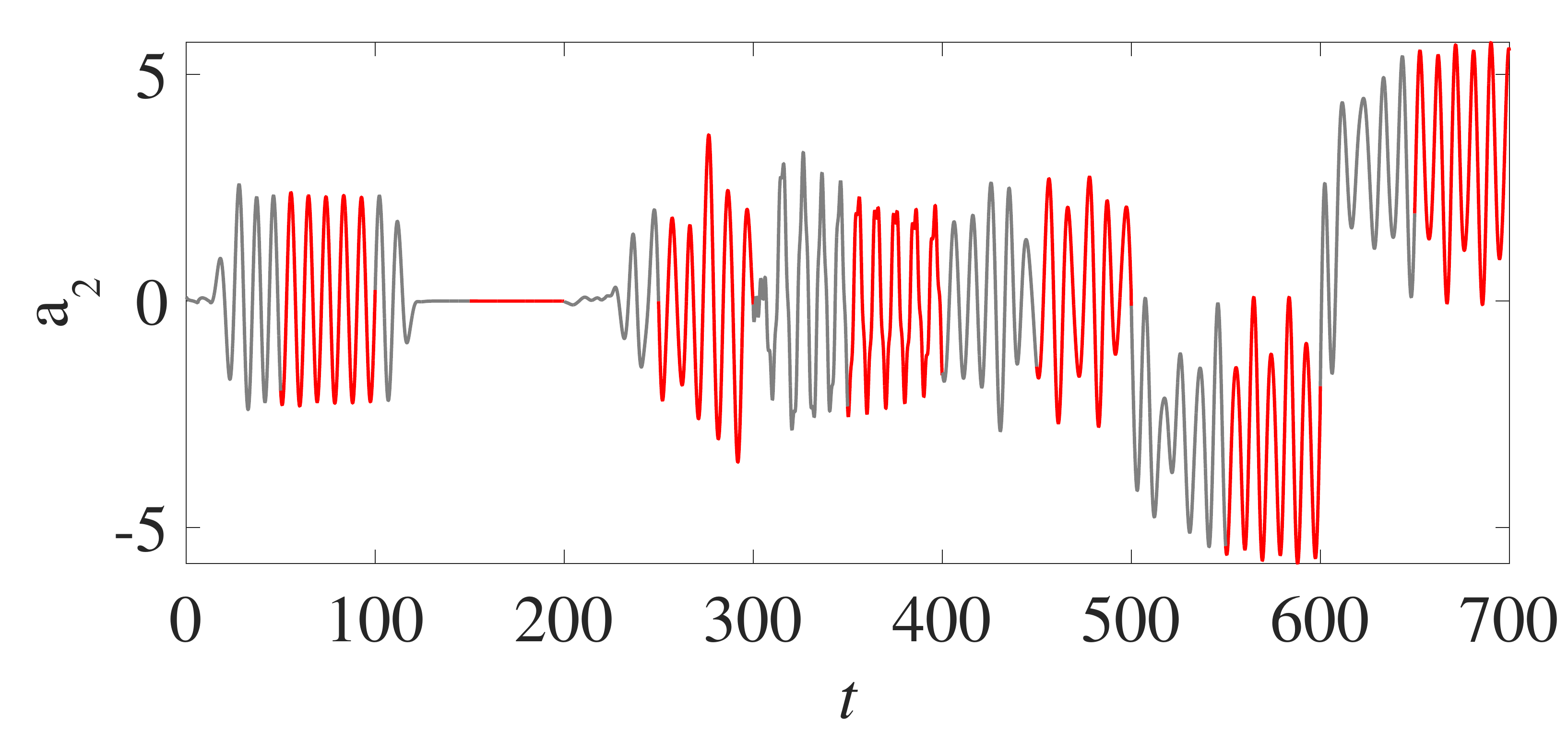}
	\includegraphics[width=6cm, trim=0 3.5cm 0 0cm, clip=true]{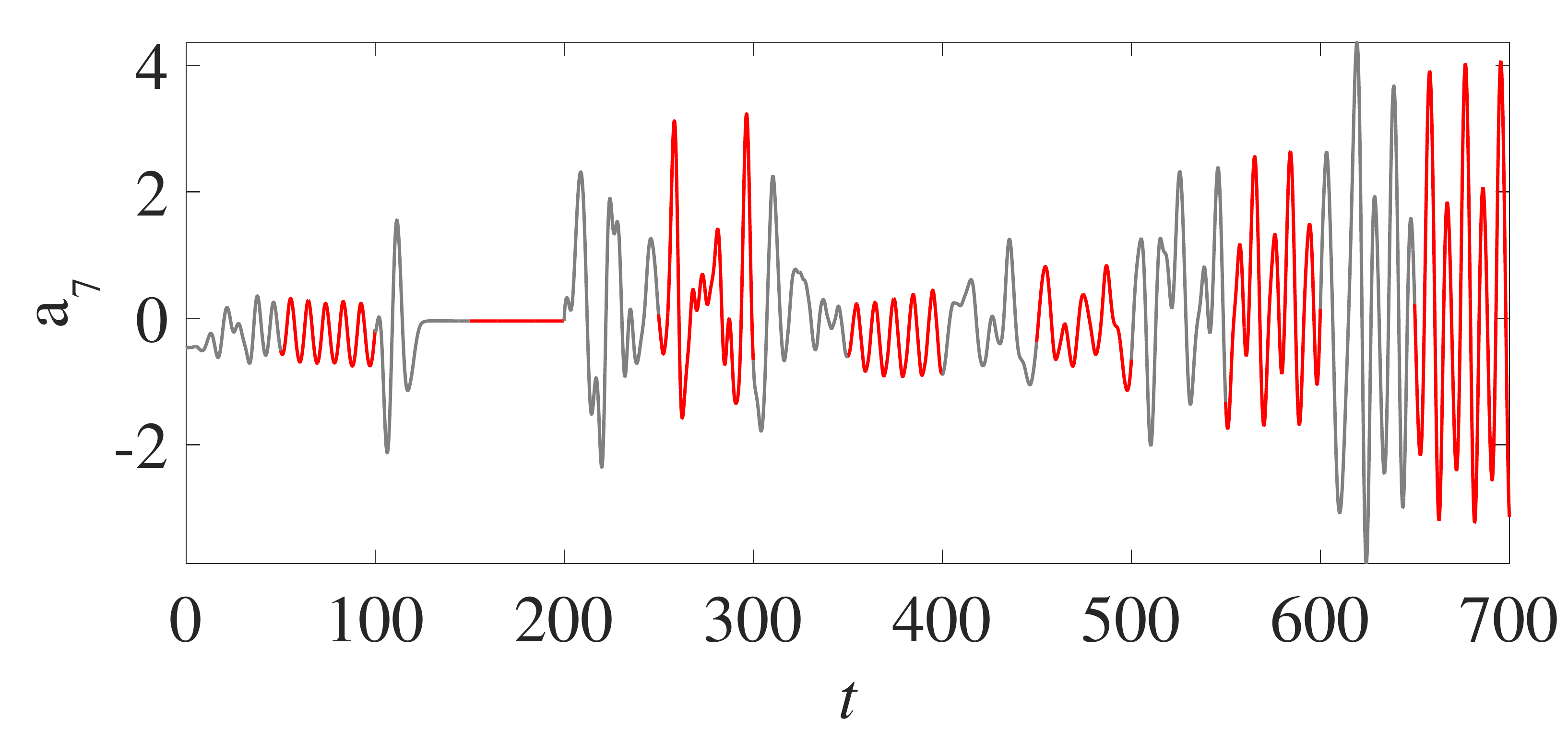}\\
	\includegraphics[width=6cm, trim=0 3.5cm 0 0cm, clip=true]{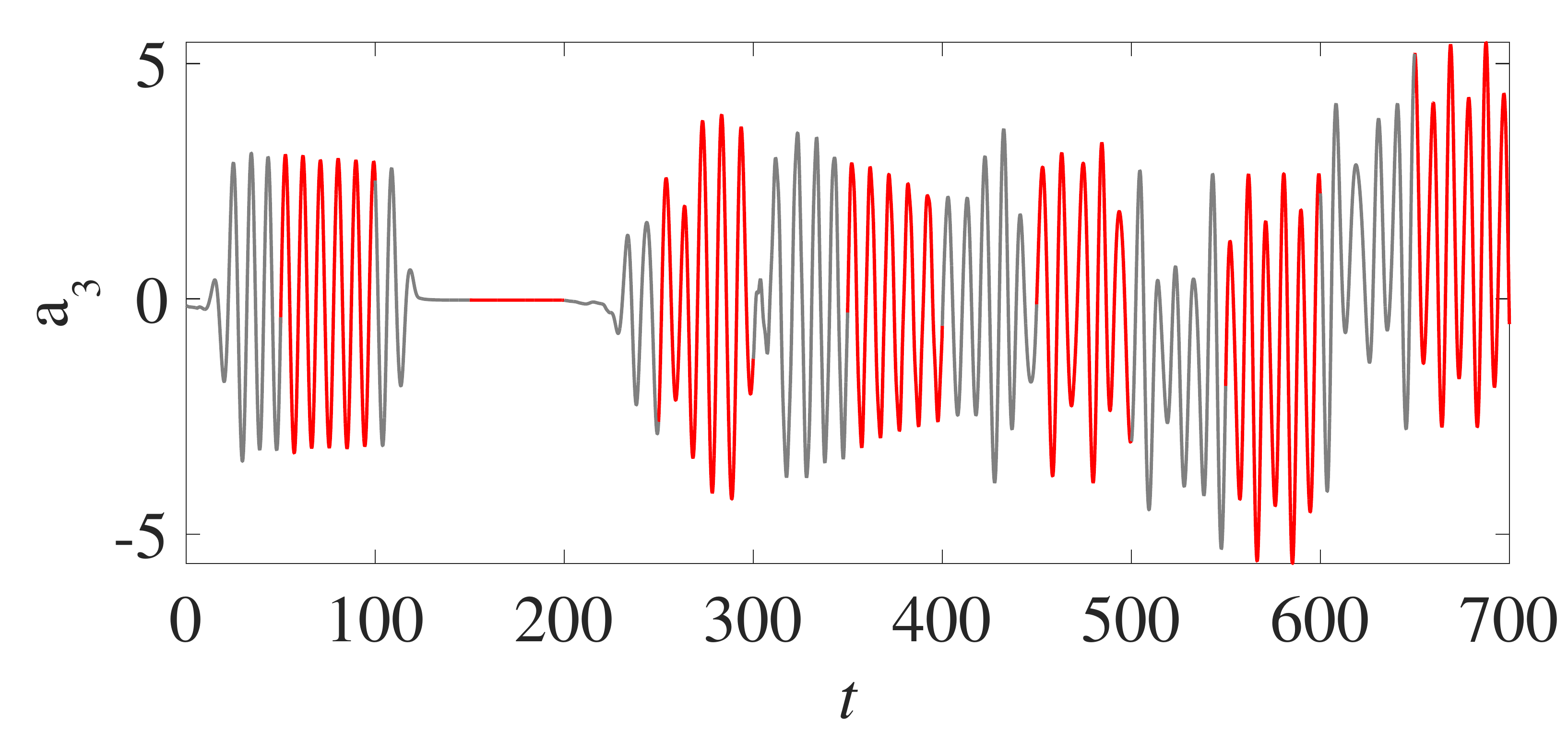}
	\includegraphics[width=6cm, trim=0 3.5cm 0 0cm, clip=true]{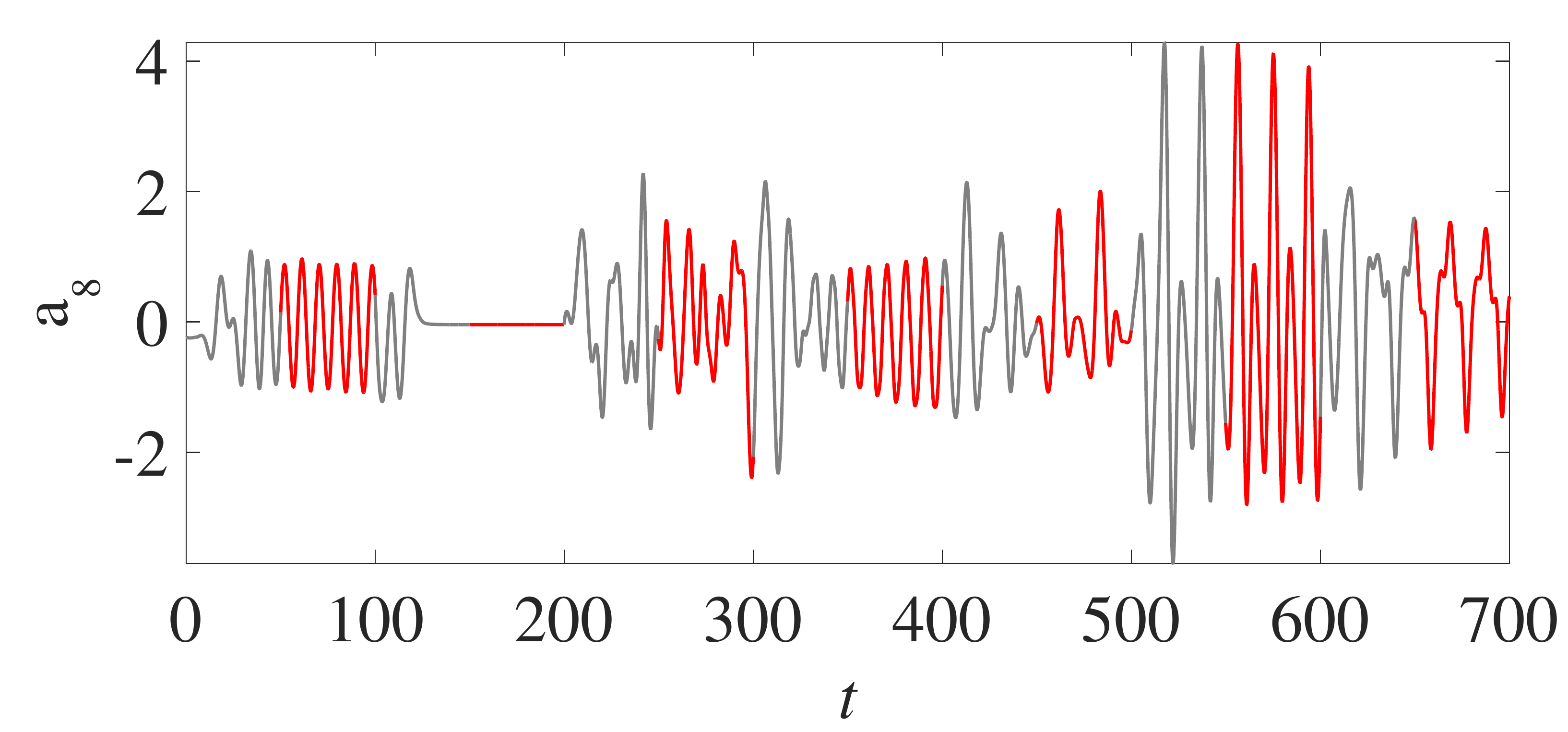}\\
	\includegraphics[width=6cm, trim=0 3.5cm 0 0cm, clip=true]{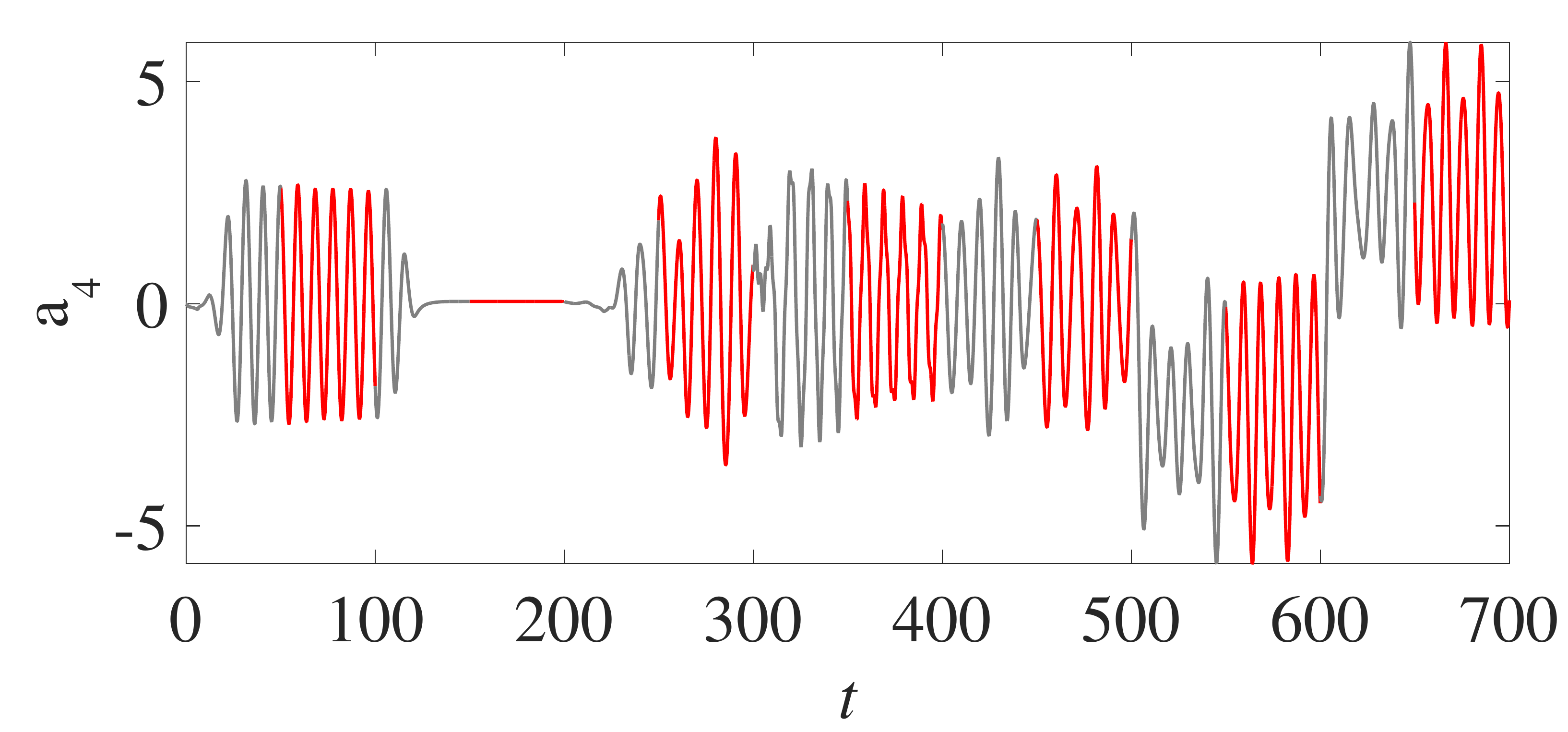}
	\includegraphics[width=6cm, trim=0 3.5cm 0 0cm, clip=true]{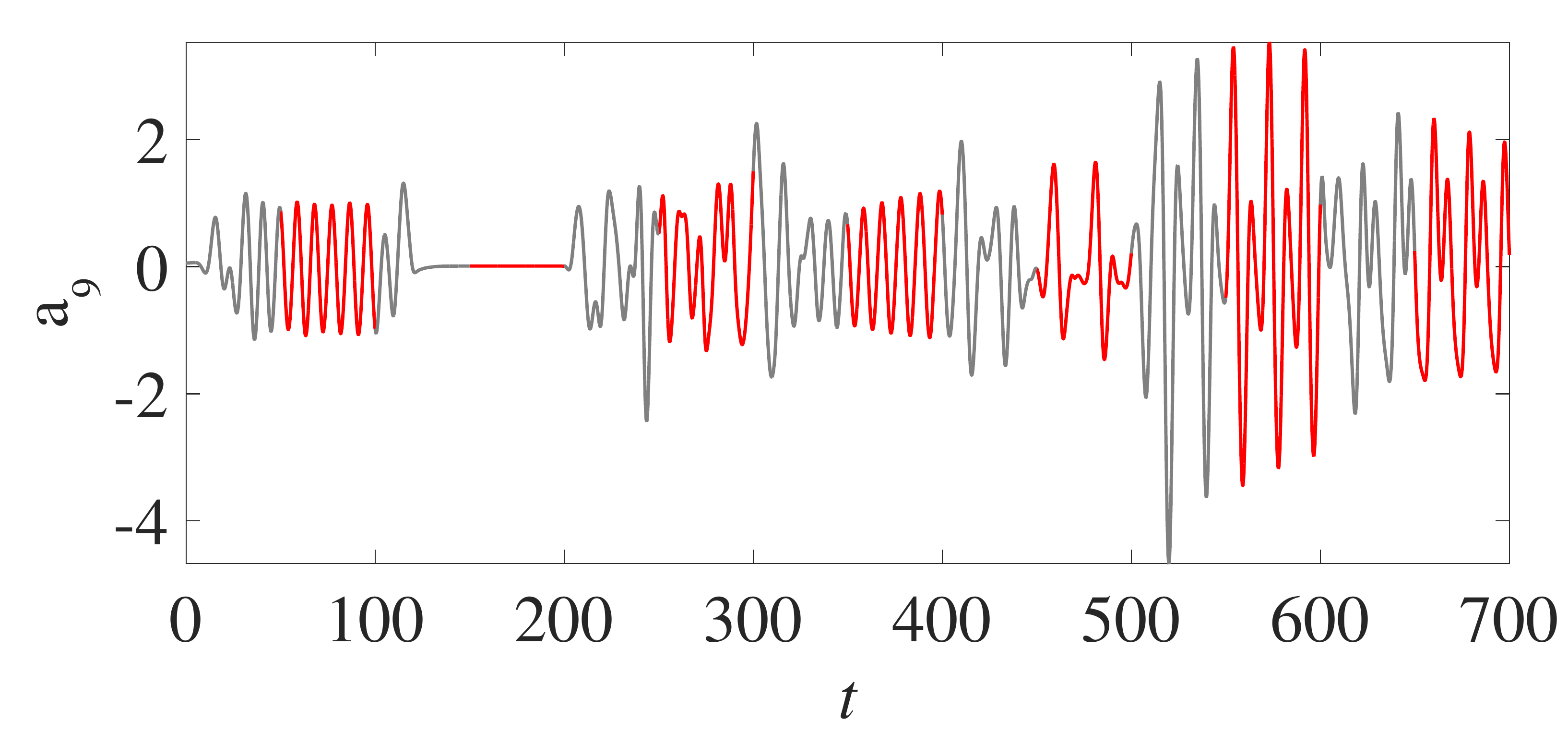}\\
	\includegraphics[width=6cm, trim=0 0cm 0 0cm, clip=true]{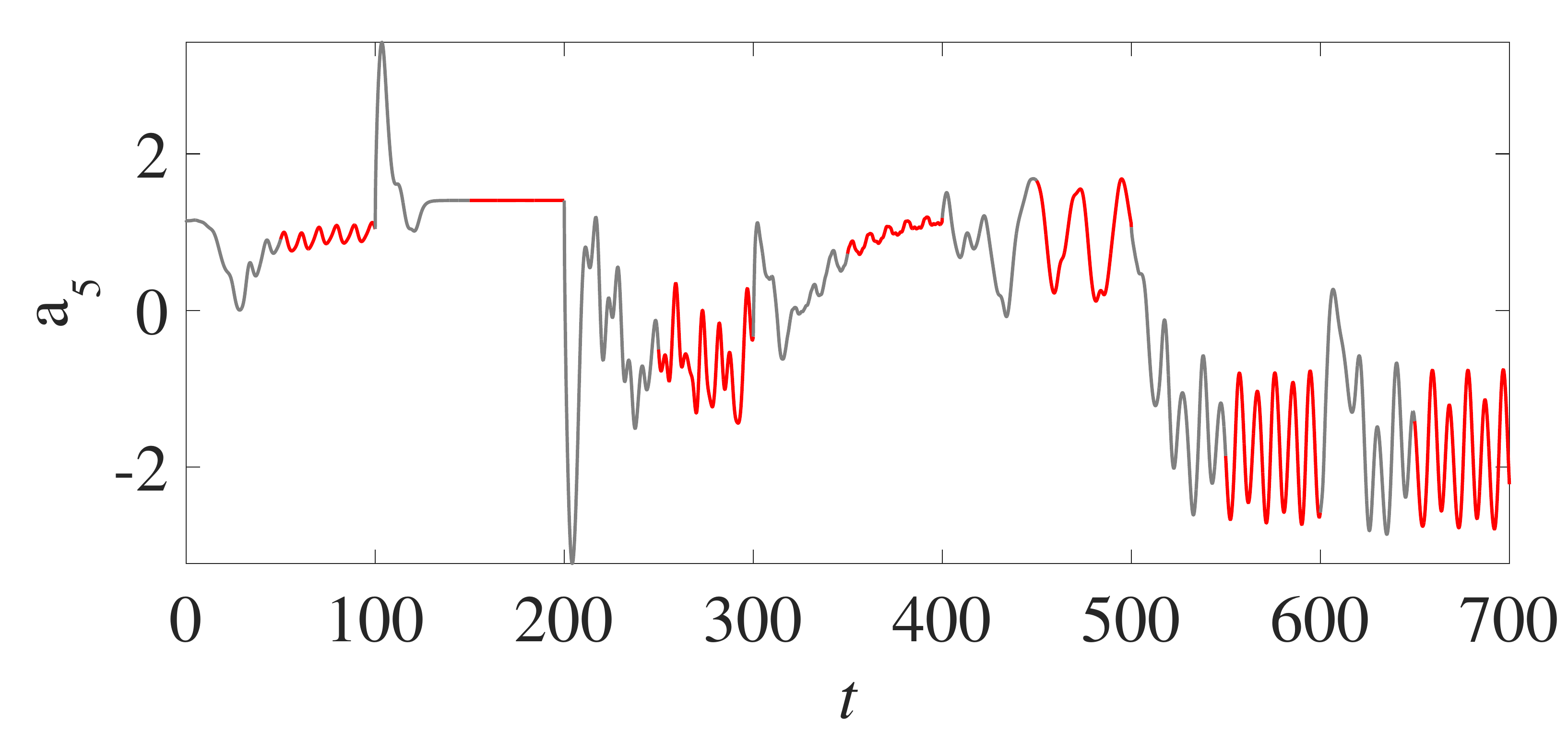}
	\includegraphics[width=6cm, trim=0 0cm 0 0cm]{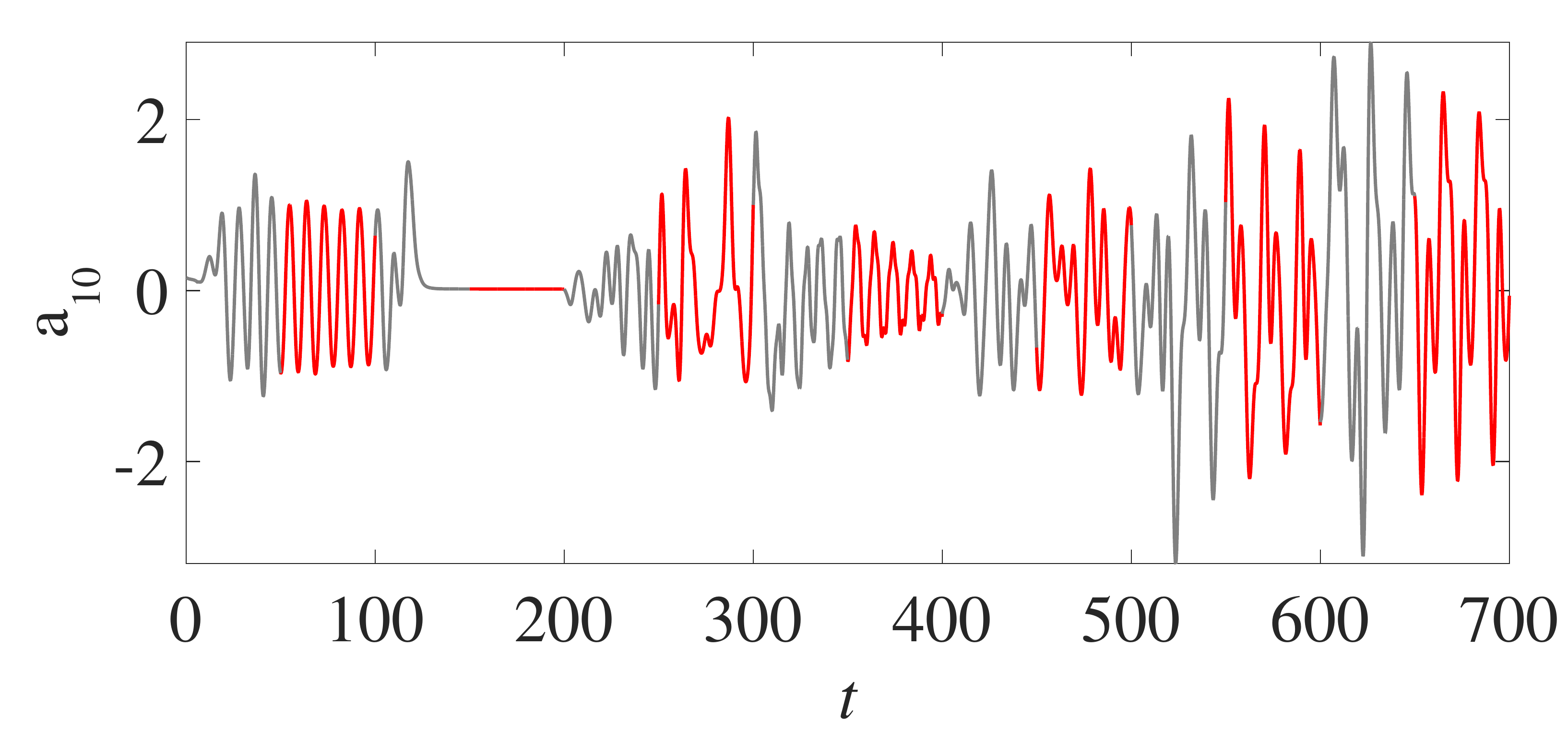}										
	\caption{POD amplitudes $a_i(t)$, $i=1,\ldots,10$. Snapshot sequences after the decay of the transients are employed for the analysis (colored in red).}
	\label{Fig:Pinball:PODAmplitudes}
\end{figure}

Figures~\ref{Fig:Pinball:PODModes} and~\ref{Fig:Pinball:PODAmplitudes} display
the first ten POD modes and their amplitudes, respectively. 
Mode 1 resolves a base-flow deformation, like a shift mode~\citep{Noack2003jfm}.
Mode 5 represents a symmetric near-field modulation.
The other modes have more oscillatory structures.
We shall not pause to hypothesize about the physical meaning of these modes.
Modes with cleaner frequency content yet near the optimal residual could,
for instance, be constructed with recursive DMD~\citep{Noack2016jfm}.
The cluster analysis resulting in the snapshot cluster affiliation (Fig.\ \ref{Fig:Pinball:SnapshotCluster})
and centroids (Fig.\ \ref{Fig:Pinball:Centroids})
is physically easier to interpret than the POD.

It may be noted that POD also allows for another metric of attractor overlap.
Let $p(\vec{a})$ be the probability distribution associated with one attractor
and $q(\vec{a})$ the distribution of another one.
Then, the continuous version of the Jensen-Shannon distance \eqref{Eqn:JensenShannonDistance}
can be applied.
An advantage is that the snapshot distance \eqref{Eqn:DistanceSnapshots}
is reflected in the construction of the metric.
A disadvantage is the need for approximations to construct a continuous probability distribution from a finite number of snapshots.

\section{Jensen-Shannon distance}
\label{Sec:JensenShannon}

Probability distributions can easily be compared using information measures.
Let $\vec{Q}=\left [Q_1, \ldots, Q_K \right ] $ 
be a reference probability distribution
and $\vec{P}=\left [P_1, \ldots, P_K \right ]$ a new measured one.
Then, the information gained from $\vec{P}$ 
with respect to the reference $\vec{Q}$ 
is quantified by the \emph{Kullback-Leibler divergence}~\citep{Kullback1951ams,Kullback1959book}, 
 also called \emph{relative entropy},
\begin{equation}
\label{Eqn:KullbackLeiblerDivergence}
D_{KL} (\vec{P} \Vert \vec{Q})
  = \sum\limits_{k=1}^K P_{k} \ln \left[ \frac{P_{k}}{Q_{k}} \right] .
\end{equation}
Identical probability distributions $\vec{P} = \vec{Q}$ 
give rise to a vanishing Kullback-Leibler divergence.
Different distributions yield a positive value.
If $P_k = 0$, the term $P_k \ln P_k/Q_k$ is interpreted as zero
because $\lim_{x \to 0} x \ln x = 0$.
Strictly speaking, the Kullback-Leibler divergence is not defined
in case there exists a $k$ for which $Q_k=0$ and $P_k>0$.
One could follow a common practice of smoothing using the \emph{absolute discounting method} by replacing
$Q_k$ by a small value, e.g.\ $\epsilon = 0.001$.

The Kullback-Leibler divergence is not symmetric,
i.e.\ $D_{KL}  ( \vec{P} \Vert \vec{Q} ) \not \equiv D_{KL} ( \vec{Q} \Vert \vec{P})$.
Hence, it cannot serve as a metric. 
The probability distributions $\vec{P}$ and $\vec{Q}$ may be compared 
with the \emph{Jensen-Shannon divergence} 
which is defined as symmetrized and smoothened Kullback-Leibler divergence:
\begin{equation}
\label{Eqn:JensenShannonDivergence}
JSD (\vec{P} , \vec{Q})
  = \frac{1}{2} \left[ D_{KL} \left( \vec{P} \Vert \vec{M} \right) 
                     + D_{KL} \left( \vec{Q} \Vert \vec{M}  \right) \right],
\quad \hbox{where} \quad \vec{M} = \frac{1}{2} \left[ \vec{P} + \vec{Q} \right].
\end{equation}
Equivalently,
\begin{equation}
\label{Eqn:JensenShannonDivergence2}
JSD (\vec{P} , \vec{Q})
  = \frac{1}{2} \sum\limits_{k=1}^K  \left( P_k \ln \left[ \frac{2 P_k}{P_k + Q_k} \right] 
                                          + Q_k \ln \left[ \frac{2 Q_k}{P_k + Q_k} \right] \right).
\end{equation}
The summation term is interpreted as zero 
if  $P_k=Q_k=0$ or if the numerator of the logarithm argument vanishes.  
Note that we don't need the $\epsilon$ threshold of the Kullback-Leibler entropy anymore.

The square root of the Jensen-Shannon divergence has all properties of a metric \citep{Endres2003ieeetit}
and is referred to as \emph{Jensen-Shannon distance},
\begin{equation}
\label{Eqn:JensenShannonDistance}
D_{JS} \left( \vec{P}, \vec{Q} \right) = \sqrt{JSD \left( \vec{P}, \vec{Q} \right)}.
\end{equation}
This distance \eqref{Eqn:JensenShannonDistance} defines
an  \emph{entropic} metric of attractor overlap 
between two attractor data 
represented by the cluster distributions $\vec{P}$ and $\vec{Q}$.

\begin{figure}
	\centering
	\begin{tabular}{ll}
		(a) & (b)  \\
	\ \ \ \ \ \includegraphics[height=4cm]{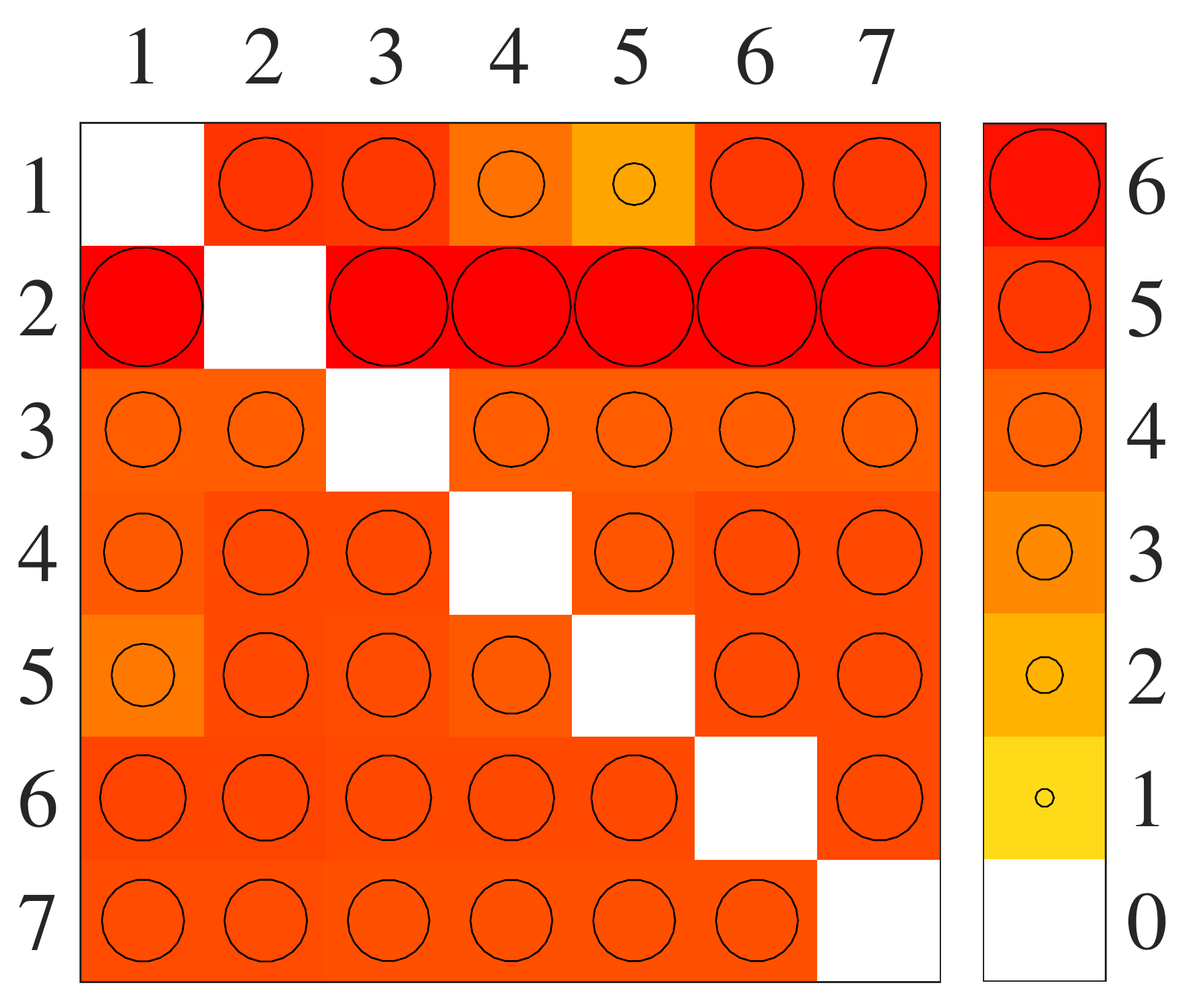} \ \ \ \ & \ \ \ \ \
	\includegraphics[height=4cm]{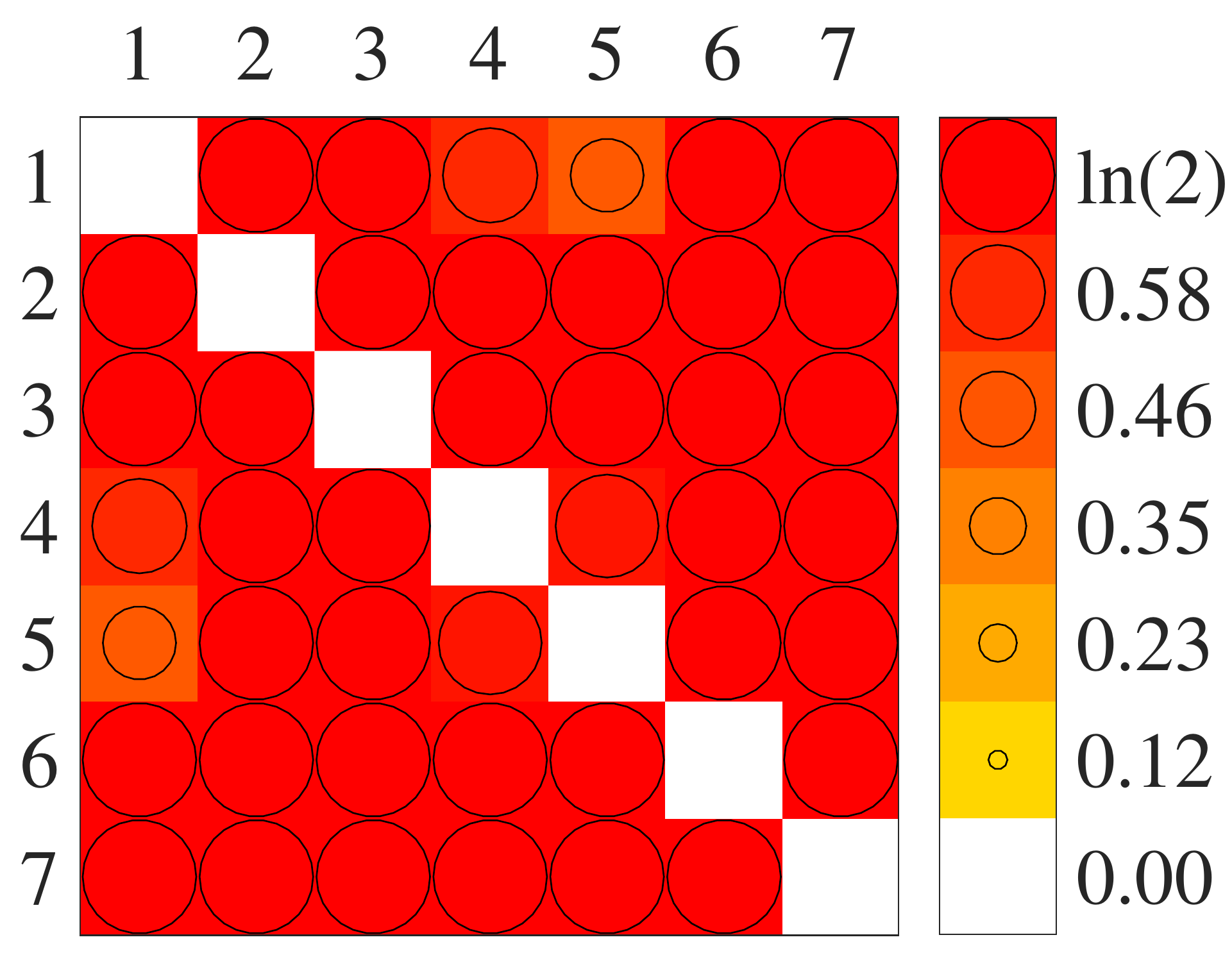}
	\end{tabular}
	\caption{Distances between attractors: 
		(a) Kullback-Leibler divergence and
		(b) Jensen-Shannon divergence
		based on the cluster probability distribution for each of the seven control phases.
                The row and column of  $D_{KL}( P \Vert Q)$ corresponds to the first ($\vec{P}$) and second argument ($\vec{Q}$).
           The value of the divergence is color-coded and indicated by the size of the circle (see the corresponding caption).
}
\label{Fig:Pinball:Entropies}
\end{figure}
Figure \ref{Fig:Pinball:Entropies} displays the Kullback-Leibler and Jensen-Shannon divergences
between the seven fluidic pinball attractors.
The diagonal entries vanish by definition.
Unforced vortex shedding is seen to be similar to high- and low-frequency forcing,
as was indicated already by the many shared clusters in  figure~\ref{Fig:Pinball:SnapshotCluster}.
The difference between unforced shedding,
base-bleed dynamics, and both Magnus effects is larger
as they share no joint clusters.
The stabilized boat tailing with a single cluster 
is seen to be very different from all other phases with no cluster overlap.
The Kullback-Leibler divergence is, as expected, not symmetric.
For instance. the divergence of boat-tailing 
with respect to base-bleed is larger than the other way round.
It should be noted that the Jensen-Shannon distance is typically $\ln(2)$ (or unity if the base 2 logarithm is used)
corresponding to no shared clusters.

The Jensen-Shannon distance is a natural metric for probability distributions.
Yet, it is blind to the geometric location of the centroids.
Let us assume the first attractor data populates only cluster 1,
$P_i = \delta_{i,1}$, and the second data occupies only cluster 2, $Q_i=\delta_{i,2}$.
In this case, $JSD(\vec{P},\vec{Q}) = \ln 2$,
regardless of whether the centroids are very close or very far from each other.
This property makes the Jensen-Shannon distance strongly dependent on the number of clusters
or, equivalently, on the typical size of the clusters.
For the minimum number of clusters, $K=1$, 
we have obtained the trivial result $P_1=Q_1=1$ and $JSD=0$.
For the maximum number of clusters,
each snapshot represents one centroid and defines one cluster.
In the generic case of different snapshots,
the Jensen-Shannon distance is also $\ln 2$ 
independent of the geometric location of the snapshots.
Only a few of the seven fluidic pinball phases share joint clusters
and the proximity maps based on \eqref{Eqn:JensenShannonDistance}
provides  limited physical insight.
The metric may be far more meaningful in other cases
in which most attractors have significant overlap.

\bibliography{Main}
\bibliographystyle{jfm}

\end{document}